\documentclass{iopart}
\usepackage[latin1]{inputenc}
\usepackage{iopams}
\usepackage{amssymb,amsfonts,graphicx,epstopdf,bm}
\usepackage[colorlinks,bookmarks,breaklinks,pdftex,urlcolor=blue,citecolor=blue,linkcolor=blue]{hyperref}
\usepackage{cite}
\newcommand{\eqref}[1]{(\ref{#1})}
\newcommand{\pd}[2]{\frac{\partial #1}{\partial #2}}
\newcommand{\pdd}[2]{\frac{\partial^{2} #1}{\partial #2 ^{2}}}
\newcommand{\bigo}{{O}}

\newcommand{\abs}[1]{\left\vert #1 \right\vert}
\newcommand{\ave}[1]{\left \langle #1 \right \rangle}
\newcommand{\explr}[1]{\exp\left[ #1 \right]}
\newcommand{\diag}{\mbox{diag}}
\def\chem#1#2{ {{\scriptscriptstyle#1\atop\longrightarrow}\atop{\longleftarrow\atop \scriptscriptstyle#2}} }
\newcommand{\hgam}{\varphi}
\newcommand{\prefac}{k}
\newcommand{\bevec}[1]{\bm{\psi}_{#1}} % evec: element form
\newcommand{\bfone}{\mathbf{1}}  % the vector  (1,1,1,...,1)^{T}
\newcommand{\blevec}[1]{\bm{\eta}_{#1}}

\newcommand{\bevecg}[1]{\tilde{\bm{\psi}}_{#1}}
\newcommand{\blevecg}[1]{\tilde{\bm{\eta}}_{#1}}
\newcommand{\befn}[1]{\bm{\phi}_{#1}}  % eigenfunctions: vector 
\newcommand{\baefn}[1]{\bm{\xi}_{#1}}
\newcommand{\aefn}[2]{\xi^{(#1)}_{#2}}
\newcommand{\qss}{\bar{\bm{\rho}}}
\newcommand{\mv}{\nu}  % mean velocity 

\eqnobysec
\begin{document}

\title{Isolating intrinsic noise sources in a stochastic genetic switch}
\author{Jay M. Newby}
\address{Mathematical Institute, University of Oxford, 24-29  St. Giles', Oxford OX1 3LB, UK}
\ead{newby@maths.ox.ac.uk}

\begin{abstract}
The stochastic mutual repressor model is analysed using perturbation methods.  This simple model of a gene circuit consists of two genes and three promotor states.  Either of the two protein products can dimerize, forming a repressor molecule that binds to the promotor of the other gene.  When the repressor is bound to a promotor, the corresponding gene is not transcribed and no protein is produced.  Either one of the promotors can be repressed at any given time or both can be unrepressed, leaving three possible promotor states.  This model is analysed in its bistable regime in which the deterministic limit exhibits two stable fixed points and an unstable saddle, and the case of small noise is considered.  On small time scales, the stochastic process fluctuates near one of the stable fixed points, and on large time scales, a metastable transition can occur, where fluctuations drive the system past the unstable saddle to the other stable fixed point.  To explore how different intrinsic noise sources affect these transitions, fluctuations in protein production and degradation are eliminated, leaving fluctuations in the promotor state as the only source of noise in the system.  Perturbation methods are then used to compute the stability landscape and the distribution of transition times, or first exit time density.  To understand how protein noise affects the system, small magnitude fluctuations are added back into the process, and the stability landscape is compared to that of the process without protein noise.  It is found that significant differences in the random process emerge in the presence of protein noise.
\end{abstract}

\section{Introduction}

Random molecular interactions can have profound effects on gene expression.  Because the expression of a gene can be regulated by a single promotor, and because the number of mRNA copies and protein molecules is often small, deterministic models of gene expression can miss important behaviors.  A deterministic model might show multiple possible stable behaviors, any of which can be realized depending on the initial conditions of the system.  Different stable behavior that depend on initial conditions allows for variability in response and adaptation to environmental conditions \cite{kaern05a}.  

Although in some cases, noise from multiple sources can push the behavior far from the deterministic model, here we focus on situation where the system fluctuates close to the deterministic trajectory (i.e., weak noise).   Of particular interest is behavior predicted by a stochastic model that is qualitatively different from its deterministic counterpart \cite{maheshri07a}, even if the fluctuations are small.  

Several interesting questions emerge when including stochastic effects in a model of gene expression.  For example, what are the different sources of fluctuations affecting a gene circuit?  Can noise be harnessed for useful purpose, and if so, what new functions can noise bring to the gene-regulation toolbox?  One way in which noise can induce qualitatively different behavior occurs when a rare sequence of random events pushes the system far enough away from one of the stable deterministic behaviors that the system transitions toward a different stable dynamic behavior, one that would never be realized in the deterministic model without changing the initial conditions.  For example, if the deterministic model is bistable, fluctuations can cause the protein concentration to shift between the different metastable protein concentrations.  This happens when fluctuations push the system past the unstable fixed point that separates two stable fixed points.

While often times a spontaneous change in gene expression might be harmfull, it might also be beneficial.  For example, in certain types of bacteria, a few individuals within a population enter a slow-growth state in order to resist exposure to antibiotics.  In a developing organism, a population of differentiating cells might first randomly choose between two or more expression profiles during their development and then later segregate into distinct groups by chemotaxis.  In both examples, switching between metastable states leads to mixed populations of phenotypic expression \cite{eldar10a}.  This leads to the question of how cells coordinate and regulate different sources of biochemical fluctuations, or noise, to function within a genetic circuit.

In many cases, the genes within a given circuit are turned on and off by regulator proteins, which are often the gene products of the circuit.  If a gene is switched on, its DNA is transcribed into one or more mRNA copies, which are in turn translated into large numbers of proteins.  Typically, the protein products form complexes with each other or with other proteins that bind to regulatory DNA sequences, or operators, to alter the expression state of a gene.  For example, a repressor binds to an operator which blocks the promotor---the region of DNA that a polymerase protein binds to before transcribing the gene---so that the gene is turned off and no mRNA are transcribed.    This feedback enables a cell to regulate gene expression, and often multiple genes interact within groups to form gene circuits. 

Understanding how different noise sources affect the behavior of a gene circuit and comparing this with how the circuit behaves with multiple noise sources is essential for understanding how a cell can use different sources of noise productively.  Fluctuations arising from the biochemical reactions involving the DNA, mRNA, and proteins are commonly classified as ``intrinsic'' noise \cite{thattai01a}.  One important source of intrinsic noise is fluctuations from mRNA transcription, protein translation, and degradation of both mRNA and protein product.   This type of noise is common among many of the biochemical reactions within a cell, and its effect is reduced as the number of reacting species within a given volume grows large.  Another source of intrinsic noise is in the expression state of the genes within the circuit.  Typically there is only one or two copies of a gene within a cell, which means that thermal fluctuations within reactions with regulatory proteins have a significant effect on mRNA production.  Here, we consider the situation where transitions in the behavior of a gene circuit are primarily driven by fluctuations in the on/off state of its promotor and examine the effect of removing all other sources of noise.  

Stochastic gene circuits are typically modelled using a discrete Markov process, which tracks the random number of mRNA and/or proteins along with the state of one or more promotors \cite{mcadams97a,aurell02a,roma05a,paulsson05a,hornos05a} (but see also \cite{friedman06b}).  Monte-Carlo simulations using the Gillespie algorithm can be used to generate exact realizations of the random process.  The process can also be described by its probability density function, which satisfies a system of linear ordinary differential equations known as the Master equation.  The dimension of the Master equation is the number of possible states the system can occupy, which can be quite large, leading to the problem of dimensionality when analyzing the Master equation directly.  However, for the problem considered here, the full solution to the Master equation is not necessary in order to understand metastable transitions.

The motivating biological question we consider here is what percentage of a population of cells can be expected to exhibit a metastable transition within a given timeframe.  If a spontaneous transition is harmfull to the cell, one expects that reaction rates and protein/DNA interactions should evolve so that transition times are likely to be much larger than the lifetime of the cell.  On the other hand, if spontaneous transition are functional, transition times should be tuned to achieve the desired population in which the transition occurs.  In either case, the key quantity of interest is the distribution of transition times between metastable states, regardless of the noise source driving the transition.  

Except for a few special cases \cite{hornos05a,shahrezaei08a} exact results, even for the mean transition time, are not possible and approximation techniques or Monte-Carlo simulations must be used.  However, because rare events typically involve long simulation times where large numbers of jumps occur, Monte-Carlo simulations are computationally expensive to perform, leaving perturbation analysis ideally suited for the task.  

Past studies of metastable transitions, where perturbation methods are applied to the Master equation, have used a simplifying assumption so that the state of the promotor is not accounted for explicitly \cite{aurell02a,roma05a}.  The assumption is that proteins are produced in ``bursts'' during which one or more mRNA copies are translated to rapidly produce many proteins.  In these models, production bursts occur as instantaneous jumps, with a predefined distribution determining the number of proteins produced during a given burst.  More recently, Assaf and coworkers analyzed a model where the on/off state of a single stochastic promotor is accounted for explicitly, and mRNA copies are produced stochastically at a certain rate when the promotor is turned on \cite{assaf11a}.  However, the case where the model contains an arbitrary number of promotors or promotor states has not been addressed, and as we show in this paper, accounting for even just three promotor states is nontrivial.   Similar asymptotic methods have also been developed to study metastable transitions in continuous Markov processes \cite{ludwig75a,matkowsky83a,talkner87a,naeh90a,maier97a,schuss10a}, but cannot be applied to a discrete chemical reaction system because continuous approximations, such as the system-size expansion, of discrete Markov processes do not, in general, accurately capture transition times \cite{doering05a}.  Another source of difficulty that arises from isolating promotor state fluctuations as the only source of noise is that the resulting state space of the Markov process is both continuous and discrete.

After removing all sources of intrinsic noise except for the fluctuating promotor---by taking the thermodynamic limit---the protein levels change deterministically and continuously, and the promotor's state jumps at exponentially-distributed random times.  The random jumps in the promotor's state makes the protein levels appear random, even though they are only responding deterministically to changes in the promotor state.  Such random processes are sometimes called hybrid systems, or piecewise deterministic \cite{zeiser09a,zeiser10a}.  Here we refer to it as the quasi-deterministic (QD) process because we are taking part of the randomly fluctuating discrete state of the system (the number of protein molecules) and replacing it with a deterministically-changing continuous state. Recently, we have developed asymptotic methods for metastable transitions---similar to those applied to the discrete Master equation---for Markov processes with both discrete and continuous state spaces \cite{keener11a,newby11b}.  However, these methods do not account for two or more continuous state variables, which restricts the genetic circuit the method can analyze to one with a single protein product.

In this paper, we develop new perturbation methods so that we can study metastable transitions in genetic circuits driven by promotor fluctuations.  These methods are based on previous theory developed for one-dimensional velocity jump processes, and are generalized to account for the multiple continuous states representing the quantity of proteins produced by the genetic circuit.  They also fit within a larger framework of methods to study metastable transitions in continuous Markov processes \cite{schuss10a} and in discrete Markov processes \cite{hanggi84a,dykman94a,hinch05a,vellela07a,escudero09a,metzner09b,bressloff10b,aurell02a,roma05a,assaf11a}.  For illustration, we use a simple model known as the ``mutual repressor'' model \cite{kepler01a}, which contains two genes, two promotors, and three promotor states.  Although our example considers only three promotor states, the methods presented are general and can account for an arbitrary number of promotor states.  For a range of parameter values, the deterministic limit of the mutual repressor model is bistable, having two stable fixed points separated by an unstable saddle point.  For the stochastic model, the deterministic forces create two confining wells surrounding each stable fixed point, separated by a stability barrier along the separatrix that contains the unstable saddle.  This geometric interpretation is given by taking the logarithm of the stationary probability density function, which we refer as the ``stability landscape.''  An approximation of the first exit time density is found for the random process to escape over the stability barrier from one of the stability wells to the other.  Using this model, we seek to answer the following question.  How does the random process change when protein noise is removed, leaving the state of the promotor as the only source of randomness?  That is, are there any qualitative differences in the behavior of the system without other sources of intrinsic noise?

The paper is organized as follows.  In Section \ref{sec:model} the Mutual Repressor model is presented along with its reduction to the QD process, and then in Section \ref{sec:trt}, perturbation methods for estimating the exit time density are applied to the QD process.  For comparison, the stability landscape is also computed for the full process in Section \ref{sec:full}, which includes fluctuations in protein production/degradation.  Finally, results are presented in Section \ref{sec:results}, and the QD process is compared to the full process, using analytical/numerical approximations and  Monte-Carlo simulations.

\section{Mutual repressor model}
\label{sec:model}
The mutual repressor model \cite{kepler01a} is a hypothetical gene circuit consisting of a single promotor driving the expression of two genes: $N$ and $M$.  Each protein product can dimerize and bind to the promotor to repress the expression of the other.  When no dimer is bound to the promotor, both genes are expressed equally.  Thus, the promotor can be in one of three states: bound to a dimer of protein one $\mathcal{O}_{0}$, unbound $\mathcal{O}_{1}$, or bound to a dimer of protein two $\mathcal{O}_{2}$.  Let the number of protein product of gene $N$ and $M$ be $n$ and $m$, respectively.  It is assumed that the mRNA and protein production steps can be combined into a single protein production rate and that the dimerization reaction is fast so that it can be taken to be in quasi-steady-state.  we then have the following transition between the three promotor states
\begin{equation}
  \label{eq:1}
  \mathcal{O}_{0}\chem{\beta \kappa}{n(n-1)\kappa}\mathcal{O}_{1}\chem{m(m-1)\kappa}{\beta \kappa}\mathcal{O}_{2},
\end{equation}
where $\kappa$ is a rate and $\beta$ is a nondimensional dissociation constant.  Protein $N$ ($M$) is produced at a rate $\alpha$ while the promotor is in states $\mathcal{O}_{0,1}$ ($\mathcal{O}_{1,2}$), and both proteins are degraded at a rate $\delta$ in all three promotor states.

The probability density function $\mathrm{p}_{k}(n,m,t)$, for promotor state $k=0,1,2$ and protein numbers $n,m=0,1,2,\cdots$, satisfies the Master equation
\begin{equation}
  \label{eq:2}
  \frac{d}{dt}\mathbf{p}(n,m,t) = [\mathbb{A}+\mathbb{W}]\mathbf{p},
\end{equation}
where
\begin{equation}
  \label{eq:3}
  \mathbb{A} = \kappa\left[
  \begin{array}{ c c c}
    -\beta & n(n-1) & 0 \\
    \beta & -n(n-1)-m(m-1) & \beta \\
    0 & m(m-1) & - \beta
  \end{array}\right]
\end{equation}
is the matrix responsible for promotor state transitions. The diagonal matrix $\mathbb{W}$, responsible for changes in protein numbers, has elements
\begin{eqnarray}
  \label{eq:4}
  \mathbb{W}_{0} &=& \mathbb{D} + \alpha(\mathbb{E}_{n}^{-}-1),\\
 \mathbb{W}_{1} &=& \mathbb{D} + \alpha(\mathbb{E}_{n}^{-}+\mathbb{E}_{m}^{-}-2),\\
  \mathbb{W}_{2} &=& \mathbb{D} + \alpha(\mathbb{E}_{m}^{-}-1) ,
\end{eqnarray}
with
\begin{equation}
  \label{eq:5}
  \mathbb{D} = \delta\left[(\mathbb{E}_{n}^{+}-1)n+(\mathbb{E}_{m}^{+}-1)m\right].
\end{equation}
The shift operators $\mathbb{E}_{j}^{\pm}$ are defined according to
\begin{equation}
  \label{eq:6}
  \mathbb{E}_{j}^{\pm}f(j) = f(j\pm 1).
\end{equation}

We now introduce the nondimensional variables $t\to t\delta$, $n = x/\gamma$, and $m=y/\gamma$, where $\gamma\equiv \delta/\alpha$.  Then, the Master equation \eqref{eq:2} for the rescaled probability density, $\mathbf{p}(n,m,t)\to \bm{\rho}(x,y,t)$, becomes
\begin{equation}
  \label{eq:7}
   \frac{d}{dt}\bm{\rho}(x,y,t) = \left[  \frac{1}{\epsilon}A_{\gamma} + \frac{1}{\gamma}W \right]\bm{\rho},
\end{equation}
with dimensionless parameters $b = \frac{\beta\delta^{2}}{\alpha^{2}}$ and $\epsilon = \frac{\delta^{3}}{\kappa\alpha^{2}}$.  The matrices are given by
\begin{equation}
  \label{eq:8}
  A_{\gamma} =\left[
  \begin{array}{c c c}
   -b & x(x+\gamma) & 0 \\
    b & -x(x+\gamma)-y(y+\gamma)& b \\
    0 & y(y+\gamma) & -b
  \end{array}\right],
\end{equation}
and 
\begin{eqnarray}
  \label{eq:9}
    W &=&((e^{\partial x}-1)x + (e^{\partial y}-1)y)I\\
\nonumber
   &&\quad + \diag( e^{-\partial x} - 1,e^{-\partial x}+e^{-\partial y}- 2,e^{-\partial y}-1).
\end{eqnarray}
The operators $e^{\pm \partial x}$ and $e^{\pm \partial x}$ are defined in terms of Taylor series expansions (in small $\gamma$) which replace the shift operators $\mathbb{E}_{n,m}^{\pm}$ with
\begin{equation}
  \label{eq:10}
  e^{\pm \partial x }f(x) \equiv  \sum_{n=0}^{\infty} \frac{(\pm\gamma)^{n}}{n!}\frac{\partial^{n}f}{ \partial x^{n}} = f(x \pm \gamma).
\end{equation}

Assume that $\gamma \ll 1 $ is a small parameter, so that there is a large average number of proteins, and assume also that the parameter $\epsilon\ll 1$ is small, which reflects rapid switching between promotor states compared to the rate of protein production/degradation.  Because we have two small parameters in our system, $\gamma$ and $\epsilon$, when perusing an asymptotic solution, we must carefully consider how the limit $\epsilon\to 0$ and $\gamma\to 0$ is taken, or more practically, how large $\epsilon$ is compared to $\gamma$.  The fluctuations in the promotor state are controlled by $\epsilon$, and in the limit $\epsilon\to 0$, the transitions are infinitely fast so that the promotor behaves deterministically.  The fluctuations in protein levels is controlled by $\gamma$, and in the limit $\gamma\to 0$ the protein production/degradation behaves deterministically.  Since we are concerned primarily with rare transitions driven by promotor fluctuations and not by fluctuations in the protein production/degradation reaction, we assume that $\gamma \ll \epsilon$ (i.e., $\frac{\alpha \kappa}{\delta^{2}} \ll 1$).

\begin{figure}[htbp]
  \centering
  \includegraphics[width=10cm]{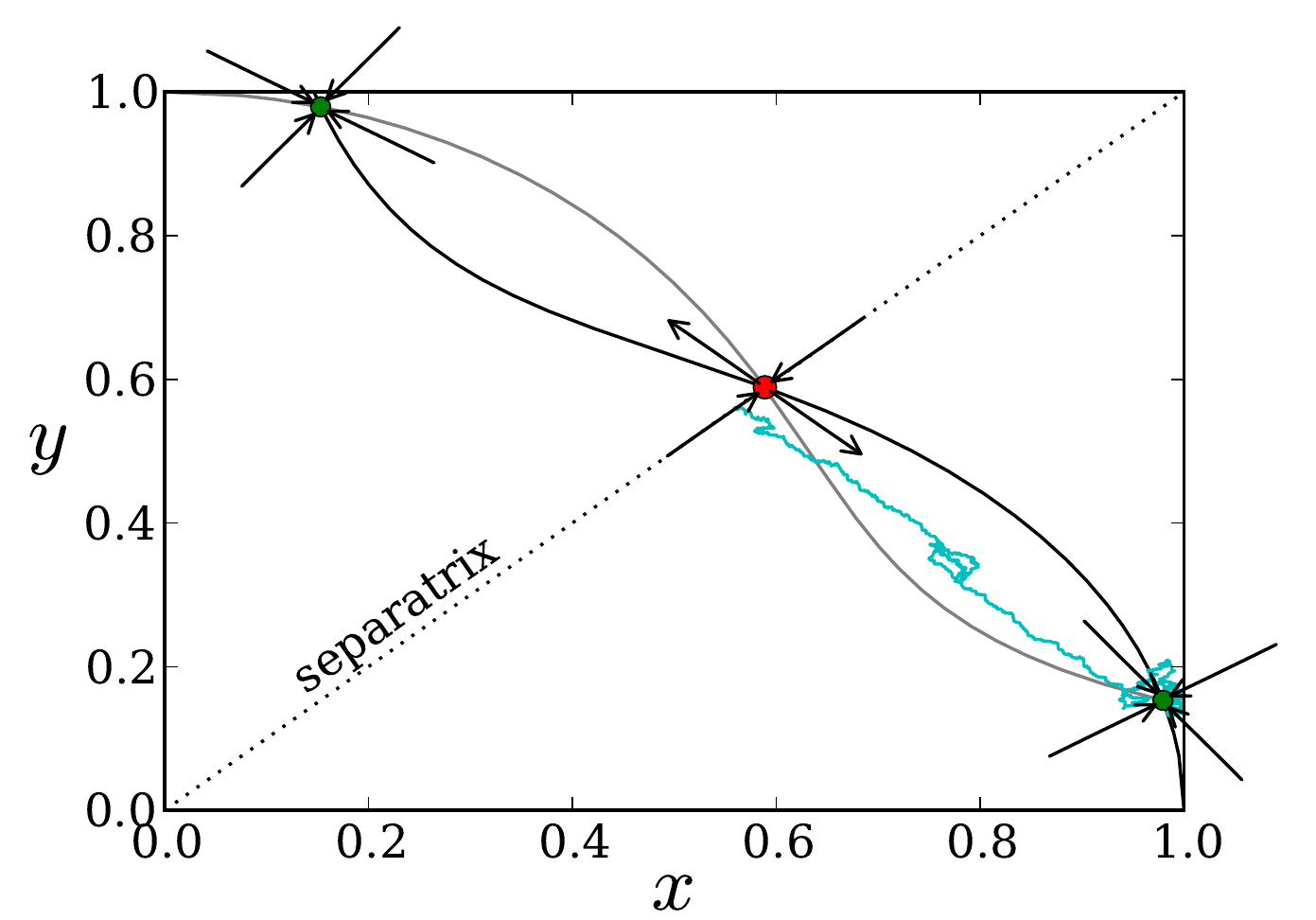}
  \caption{Deterministic dynamics for $b=0.15$.  The black curve shows the $y$-nullcline and the grey curve shows the $x$-nullcline.  The green circles show the stable fixed points, the red circle shows the unstable saddle.  The blue curve shows a stochastic trajectory leaving the lower basin of attraction to reach the separatrix.}
  \label{fig:det}
\end{figure}

Taking both limits, $\epsilon\to0$ and $\gamma\to0$, yields the fully-deterministic dynamics,
\begin{equation}
  \label{eq:11}
  \dot{x} = f(x,y),\quad \dot{y} = f(y,x),
\end{equation}
where
\begin{equation}
  \label{eq:12}
  f(x,y) \equiv \frac{1}{1+\frac{y^{2}}{b+x^{2}}}-x.
\end{equation}
Note the symmetry in the problem; the deterministic system is unchanged if we exchange $x\leftrightarrow y$.  Dynamically, the system is bistable for $0<b<b_{u}$.  At $b=b_{u}=4/9$ there is a saddle-node bifurcation, and for $b>b_{u}$ there is a single stable fixed point.  we consider the case of bistability and chose $b\ll b_{u}$.  In Fig.~\ref{fig:det} the nullclines and fixed points are shown.  
The two stable fixed points are located near the corners, and the unstable saddle point is located along the separatrix.  Arrows show the eigenvectors of the Jacobian with their direction determined by the sign of the eigenvalues, all of which are real.  A stochastic trajectory that starts at the lower stable fixed point remains nearby for a long period of time until a rare sequence of jumps carries it to the separatrix.  Because the separatrix is the stable manifold, trajectories are most likely to exit near the unstable saddle point.

To remove protein noise from the system, consider the limit $\gamma\to 0$, with $\epsilon\ll 1$ fixed, so that the protein production/degradation process is deterministic within each promotor state, while the promotor state remains random.  The master equation \eqref{eq:7} becomes
\begin{equation}
  \label{eq:13}
  \pd{\bm{\rho}}{t} = -\pd{}{x}( F\bm{\rho}) - \pd{}{y}(G\bm{\rho}) + \frac{1}{\epsilon}A\bm{\rho},\quad (x,y)\in (0,1)^{2},
\end{equation}
where 
\begin{equation}
  \label{eq:14}
  F(x,y) \equiv \diag(1-x,1-x,-x),\quad G(x,y)\equiv \diag(-y,1-y,1-y).
\end{equation}
and 
\begin{equation}
  \label{eq:15}
  A =  \left[ \begin{array}{c c c}
   -b & x^{2} & 0 \\
    b & -x^{2}-y^{2}& b \\
    0 & y^{2} & -b
  \end{array}\right].
\end{equation}

\section{Transition rate theory}
\label{sec:trt}
The focus of the remaining analysis is to obtain an accurate approximation of the first exit time density function (FETD) for the QD process to evolve from one metastable state to the another.  To obtain the FETD, we supplement an absorbing boundary condition to the governing equation along the separatrix, 
\begin{equation}
\label{eq:19}
\Gamma \equiv \{(x,y):0<x=y<1\},
\end{equation}
of the deterministic dynamics, which is the barrier the process must surmount in order to transition to the other metastable state.  For the Master equation \eqref{eq:2}, the absorbing boundary condition is simply 
\begin{equation}
  \label{eq:20}
  \mathbf{p}(n,n,t) = 0,\quad \mbox{for } n=0,1,\cdots.
\end{equation}
For the QD CK equation \eqref{eq:13}, the absorbing boundary condition is 
\begin{equation}
  \label{eq:21}
  \rho_{n}(x,y) = 0,\quad \mbox{for }(x,y)\in \Gamma,\; \mbox{and } (F_{n,n},G_{n,n})\cdot\hat{\mathbf{n}}<0,
\end{equation}
where $\hat{\mathbf{n}}$ is the unit vector normal to the boundary $\Gamma$.  Then, the domain for the QD process with the absorbing boundary is given by
\begin{equation}
  \label{eq:22}
  \mathcal{D} \equiv (0,1)^{2}\cap \{x\leq y\}.
\end{equation}
Note the choice of the lower triangular region (instead of the upper triangular region with $x\geq y$) is arbitrary due to the symmetry in the problem.

To see how the absorbing boundary on $\Gamma$ sets up the exit time problem, define $T$ to be the random time at which the separatrix is reached for the first time, given that the process starts at the stable fixed point $(x_{*},y_{*})\in\mathcal{D}$.  Consider the survival probability
\begin{equation}
  \label{eq:23}
  \mathcal{S}(t) \equiv \sum_{n=0}^{3}\int_{\mathcal{D}}\rho_{n}(x,y,t)dA,
\end{equation}
which is the probability that $t<T$.  The FETD (or probability density function for $T$) is then
\begin{equation}
  \label{eq:24}
  \mathcal{F}(t) = -\frac{d\mathcal{S}}{dt}.
\end{equation}

The FETD for the QD process can be approximated using perturbation methods as follows.  Suppose we have a CK equation of the form
\begin{equation}
  \label{eq:25}
  \partial_{t}\bm{\rho} = -\mathcal{L}_{\epsilon}\bm{\rho},
\end{equation}
where $\mathcal{L}_{\epsilon}$ is a linear operator acting on the continuous and discrete state variables of the density function.  In the case of the QD CK equation \eqref{eq:13}, we have
\begin{equation}
  \label{eq:26}
\mathcal{L}_{\epsilon} =    \pd{}{x}( F\bm{\rho}) + \pd{}{y}(G\bm{\rho}) - \frac{1}{\epsilon}A\bm{\rho}, \quad (x,y)\in \mathcal{D}
\end{equation}
Assume that $\mathcal{L}_{\epsilon}$ has a complete set of eigenfunctions, $\{\befn{j}\}$, so that the solution can be written
\begin{equation}
  \label{eq:27}
  \bm{\rho}(x,y,t) = \sum_{j=0}^{\infty}c_{j}\befn{j}(x,y)e^{-\lambda_{j}t},
\end{equation}
for some constants $c_{j}$, and that all of the eigenvalues, $\lambda_{j}$, are nonnegative.  Assume further that if we impose a reflecting boundary condition on $\Gamma$ then the principal eigenvalue, $\lambda_{0}=0$, is the only zero eigenvalue and the eigenfunction, $\befn{0}$, is the stationary density of the process (after appropriate normalization).  Furthermore, we assume that the stationary density is exponentially small on the boundary.  Then, if instead we place an absorbing boundary condition on $\Gamma$, the stationary density no longer exists and the principal eigenvalue is exponentially small in $\epsilon$, while the remaining eigenvalues are much larger so that the solution resembles the stationary density after some initial transients.  It is this difference in time scales that we exploit to approximate the FETD.

A universal feature of the FETD for rare events is its exponential form (which follows from the separation of time scales) because the time dependence is $e^{-\lambda_{0}t}$, for $\lambda_{1}t\gg 1$.  Indeed the FETD \eqref{eq:24} is 
\begin{equation}
  \label{eq:120}
  \mathcal{F}(t) \sim \lambda_{0}e^{-\lambda_{0}t},\quad\mbox{for }\lambda_{1}t\gg1.
\end{equation}
Thus, for large times the FETD is completely characterized by the principal eigenvalue, $\lambda_{0}$. The mean exit time is simply $1/\lambda_{0}$, which means that the eigenvalue also has the physical interpretation of the rate at which metastable transitions occur.  To obtain an approximation of this eigenvalue, we use a spectral projection method, which makes use of the adjoint operator $\mathcal{L}_{\epsilon}^{*}$.  Consider the adjoint eigenfunctions, $\{\baefn{j}\}$, $j=0,1,\cdots$, satisfying
\begin{equation}
  \label{eq:28}
  \mathcal{L}_{\epsilon}^{*}\baefn{j} = \lambda \baefn{j},
\end{equation}
and $\ave{\befn{i},\baefn{j}} = \delta_{i,j}$ so that the two sets of eigenfunctions are biorthogonal.  

Soppose that the boundary condition is ignored and $\befn{0}$ is approximated by the stationary density, $\qss$.  By application of the divergence theorem, the adjoint operator is such that
\begin{equation}
  \label{eq:29}
  \ave{\qss,\mathcal{L}_{\epsilon}^{*}\baefn{0}} = \ave{\mathcal{L}_{\epsilon} \qss,\baefn{0}}+\oint_{\Gamma}\baefn{0}^{T}(F-G)\qss\; ds 
\end{equation}
where the boundary contribution is nonzero because $\qss$ does not satisfy the absorbing boundary condition.  Then, since $\mathcal{L}_{\epsilon}\qss=0$, the principal eigenvalue is
\begin{equation}  
   \label{eq:30}
  \lambda_{0} = \frac{\oint_{\Gamma}\baefn{0}^{T}(F-G)\qss\; ds}{\int_{\mathcal{D}} \baefn{0}^{T}\qss \: dA}.
\end{equation}
In the remainder of this section, we use \eqref{eq:30} to approximate the principle eigenvalue by approximating the stationary density, $\qss$, and the adjoint eigenfunction, $\baefn{0}$.

\subsection{Quasi-stationary distribution}
\label{sec:qsd}
In this section, we obtain an approximation to the quasi-stationary density, $\qss(x,y)$, using a WKB approximation method.  we begin by illustrating the procedure for the QD process.  That is, we seek an approximation of the solution to the equation
\begin{equation}
  \label{eq:31}
   \left[  \frac{1}{\epsilon}A - \pd{}{x}F - \pd{}{y}G \right]\qss(x,y) = 0.
\end{equation}
Consider the anzatz
\begin{equation}
  \label{eq:32}
  \qss(x,y) = (\mathbf{r}_{0}(x,y)+\epsilon\mathbf{r}_{1}(x,y))\explr{\frac{1}{\epsilon}\Phi(x,y) + \prefac(x,y)},
\end{equation}
where $\mathbf{r}_{0,1}(x,y)$ are $3$-vectors and both $\Phi(x,y)$ and $\prefac(x,y)$ are scalar functions.  Note that in other studies of gene regulation models where similar methods are used, the small parameter in the exponential is $\gamma$.  This difference in scaling arrises from the assumption that the metastable transitions are driven by fluctuation in the promotor and not the production of protein.  Substituting \eqref{eq:32} into \eqref{eq:31} and collecting leading order terms in $\epsilon$ yields
\begin{equation}
  \label{eq:33}
  [A + pF +qG]\mathbf{r}_{0} = 0,
\end{equation}
where
\begin{equation}
  \label{eq:34}
  p\equiv \pd{\Phi}{x},\quad q \equiv \pd{\Phi}{y}.
\end{equation}

The prefactor term $\mathbf{r}_{0}$ (up to a normalization factor) is simply the nullspace of the matrix $M=[A+pF+qG]$, and we assume that it is normalized so that $\sum_{n=0}^{2}(r_{0})_{n} = 1$.  Using Theorem 3.1 in Ref.~\cite{newby11b}, we can provide necessary and sufficient conditions for $\mathbf{r}_{0}$ to be unique and positive.  For any fixed $(x,y,p,q)$, there exists a unique vector $\mathbf{r}_{0}>0$, satisfying \eqref{eq:33} if and only if the diagonal matrix $H\equiv pF+qG$ is such that at least two of its elements have oposite sign.  That is, there exist $i,j$, with $i\neq j$, such that $H_{i,i}H_{j,j}<0$.  It is interesting to speculate that once the solution $(p(x,y),q(x,y))$ to \eqref{eq:61} is substituted into the matrix $M$ that this requirement is satisfies for all $(x,y)\in (0,\infty)^{2}$.  However, this is not necessarily the case, which means that the quasi-stationary density is restricted to a subdomain where $\mathbf{r}_{0}(x,y)>0$.  

It is obvious that the protein levels must be bounded within the domain $(x,y)\in (0,1)^{2}$ when protein production/degradation is deterministic.  That is, if the gene remains in the unrepressed state, the protein level tends toward the mean value ($n,m=1/\gamma$ or $x,y=1$) but never exceeds it since protein levels do not fluctuate (unless the promotor state fluctuates).  However, it is not as obvious that the total amount of protein is further bounded so that $1<x+y<2$, which means the domain is further restricted to the upper triangular portion of the unit square.  This means that once a trajectory enters, it remains in this domain for all time and cannot escape.  To show this, we need simply look at the rate of protein production/degradation for each state normal to the line $y=1-x$.  This gives the rate for each of the promotor states ($x$ on, both on, $y$ on) when both protein levels satisfy $y=1-x$.  These rates are given by the diagonal components of the matrix $ F(x) + G(1-x) = \diag(0,1,0)$, where $F$ and $G$ are defined in \eqref{eq:14}.  It is evident that when no repressor is bound and both proteins are produced, the flux across this line is in the positive direction, and when one repressor is bound, there is no flux across this line.  

The leading order result \eqref{eq:33} defines a nonlinear partial differential equation for $\Phi$,
\begin{equation}
  \label{eq:36}
  \mathcal{H}(x,y,p,q) \equiv \mbox{det}[A(x,y)+pF(x,y) +qG(x,y)] = 0.
\end{equation}
% where
% \begin{equation}
%   \label{eq:118}
%   \begin{split}
%   \mathcal{H}(x,y,p,q) &= -x(x - 1)^{2} p^3 - y (y - 1)^2 q^3 \\
% &\qquad + (x-1) (2 x + y - 3 x y - 1) p^2 q  + (y - 1) (x + 2 y - 3 x y - 1) p q^2 \\
% &\qquad- (x - 1) (x^3 + x y^2 + 2 b x - b) p^2- (y - 1) (x^2 y + y^3 + 2 b y - b) q^2\\
% &\qquad- (2 x^3 y + 2 x y^3 - x^3- y^3 - x^2 y - x y^2 + x^2 + y^2  - 3 b (x + y ) + 4 b x y + 2 b) p q \\
% &\qquad - b (x^3 - x^2 + x y^2 + b x - b) p  - b (x^2 y + y^3 - y^2 + b y - b) q.
% \end{split}
% \end{equation}
The function $\mathcal{H}$ is referred to as the Hamiltonian for the system, due to the similarity to classical hamiltonian dynamics.  An implicit assumption,
\begin{equation}
   \label{eq:37}
  \pd{p}{y} - \pd{q}{x} = 0,
\end{equation}
 is present to ensure that $(p,q)$ is the gradient of the scalar field $\Phi$.  The above system can be solved by the method of characteristics to obtain the system of ordinary differential equations (see \cite{ockendon03a} pg.~360),
\begin{eqnarray}
  \label{eq:38}
  \dot{x} &=& \pd{\mathcal{H}}{p}, \quad \dot{y} = \pd{\mathcal{H}}{q},\\
  \dot{p} &=&-\pd{\mathcal{H}}{x}, \quad \dot{q} = -\pd{\mathcal{H}}{y},
\end{eqnarray}
where each variable is parameterized by $t$, which should not be confused with physical time.  The above system of ordinary differential equations is supplemented with an equation for the stability lanscape
\begin{eqnarray}
  \label{eq:39}
  \dot{\Phi}_{0} &\equiv& \pd{\Phi}{x}\dot{x} + \pd{\Phi}{y}\dot{y} \\
   &=& p\pd{\mathcal{H}}{p} + q \pd{\mathcal{H}}{q}. 
\end{eqnarray}
and solutions specify $\Phi$ along a curve in the plane $(x(t),y(t))$.  A family of curves is defined by specifying Cauchy data
\begin{equation}
\label{eq:126}
x(0) = x_{0}(\theta),\quad y(0)=y_{0}(\theta), \quad p(0) = p_{0}(\theta),\quad q(0)=q_{0}(\theta),
\end{equation}
along a curve parameterized by $\theta$.  

One of the difficulties found in this method is determining Cauchy data.  At the stable fixed points, the value of each of the variables is known (i.e., $p=q=0$ and $x=x_{*},y=y_{*}$) but data at a single point cannot hope to generate a family of rays.  Therefore, data must be specified on an ellipse surrounding the fixed point, using the Hessian matrix,
\begin{equation}
  \label{eq:40}
  Z \equiv \left[
  \begin{array}{c c}
    \frac{\partial^{2}\Phi}{\partial x^{2}} &  \frac{\partial^{2}\Phi}{\partial x \partial y}\\
    \frac{\partial^{2}\Phi}{\partial y \partial x} &  \frac{\partial^{2}\Phi}{\partial y^{2}}
  \end{array}\right].
\end{equation}
Expanding the function $\Phi$ in a Taylor series around the fixed point yields the quadratic form, 
\begin{equation}
  \label{eq:41}
  \Phi(x,y) \approx \frac{1}{2}r^{T}Zr,\quad r = 
\left(  \begin{array}{c}
    x-x_{*}\\
    y-y_{*}
  \end{array}\right),
\end{equation}
as its leading order term.  Cauchy data is specified on the ellipse 
\begin{equation}
  \label{eq:42}
  \frac{1}{2}r(\theta)Zr(\theta)^{T} = \omega,
\end{equation}
for some suitably small $\omega\ll 1$.  In practice, $\omega$ must be small enough to generate accurate numerical results, but large enough so that trajectories can be generated to cover the domain.  On the elliptical contour, the initial values for $p,q$ are
\begin{equation}
  \label{eq:43}
  \left(  \begin{array}{c}
    p_{0}(\theta) \\
    q_{0}(\theta)
  \end{array}\right) = Z
\left(\begin{array}{c}
  x_{0}(\theta)-x_{*} \\
  y_{0}(\theta)-y_{*}
\end{array}\right)
.
\end{equation}

It can be shown \cite{schuss10a} that the Hessian matrix is the solution to the algebraic Riccati equation,
\begin{equation}
  \label{eq:44}
  ZBZ + ZC + C^{T}Z = 0,
\end{equation}
where
\begin{equation}
  \label{eq:45}
  B = \left[
    \begin{array}{c c}
      \frac{\partial^{2} \mathcal{H}}{\partial p^{2}} & \frac{\partial^{2} \mathcal{H}}{\partial p \partial q} \\
      \frac{\partial^{2} \mathcal{H}}{\partial q \partial p} & \frac{\partial^{2} \mathcal{H}}{\partial q^{2}}
    \end{array}\right],\quad
C = \left[
  \begin{array}{c c}
    \frac{\partial^{2} \mathcal{H}}{\partial p \partial x} &  \frac{\partial^{2} \mathcal{H}}{\partial p \partial y} \\
    \frac{\partial^{2} \mathcal{H}}{\partial q \partial x} &  \frac{\partial^{2} \mathcal{H}}{\partial q \partial y}
  \end{array}\right],
\end{equation}
evaluated at $p=q=0$, $x=x_{*}$, and $y=y_{*}$.  This equation can be transformed into a linear problem (in order to actually solve it) by making the substitution $Q=Z^{-1}$ to get
\begin{equation}
  \label{eq:46}
  B + CQ + QC^{T} = 0.
\end{equation}

\subsubsection{An equation for $\prefac(x,y)$}
\label{sec:prefac}
An equation for the scalar function $\prefac(x,y)$ is found by substituting \eqref{eq:32} into \eqref{eq:31} and keeping second order terms in $\epsilon$, to get
\begin{equation}
  \label{eq:47}
\left  [A + pF + qG\right] \mathbf{r}_{1} = \pd{}{x}(F\mathbf{r}_{0}) + \pd{}{y}(G\mathbf{r}_{0}) - \left(\pd{\prefac}{x}F - \pd{\prefac}{y}G\right)\mathbf{r}_{0}.
\end{equation}
For solutions $\mathbf{r}_{1}$ to exist, the Fredholm Alternative Theorem requires that for
\begin{equation}
  \label{eq:48}
  \bm{l}^{T}[A+pF+qG] = 0
\end{equation}
 it must be that
\begin{equation}
  \label{eq:49}
  \bm{l}^{T}\left[ \pd{}{x}(F\mathbf{r}_{0}) + \pd{}{y}(G\mathbf{r}_{0}) - \left(\pd{\prefac}{x}F - \pd{\prefac}{y}G\right)\mathbf{r}_{0}  \right] = 0.
\end{equation}
Note that if $\mathbf{r}_{0}$ spans the right nullspace of $A+pF+qG$, then the left nullspace is also one dimensional.  After rewriting \eqref{eq:49}, we have the PDE for $\prefac$ given by
\begin{equation}
  \label{eq:50}
  \pd{\prefac}{x}(\bm{l}^{T}F\mathbf{r}_{0}) + \pd{\prefac}{y}(\bm{l}^{T}G\mathbf{r}_{0}) = \bm{l}^{T}\left( \pd{}{x}(F\mathbf{r}_{0}) + \pd{}{y}(G\mathbf{r}_{0})\right).
\end{equation}
Although the solution to this equation can be formulated by the method of characteristics, it requires values of the vectors $\mathbf{r}_{0}$ and $\bm{l}$, which in turn require the solution to the ray equations \eqref{eq:38}.  Since rays must be integrated numerically in most cases, solving \eqref{eq:50} along its own characteristics is impractical.  However, \eqref{eq:50} can be computed along the characteristic curves of \eqref{eq:38} as follows.  First, differentiating \eqref{eq:50} along characteristics yields
\begin{equation}
  \label{eq:51}
  \dot{\prefac} = \pd{\prefac}{x}\dot{x} + \pd{\prefac}{y} \dot{y}  = \pd{\prefac}{x}\pd{\mathcal{H}}{p} + \pd{\prefac}{y}\pd{\mathcal{H}}{q}.
\end{equation}
Using the fact that $(\bm{l}^{T}G\mathbf{r}_{0})\dot{x} - (\bm{l}^{T}F\mathbf{r}_{0})\dot{y} = 0$ along characteristics, we can define
\begin{equation}
  \label{eq:122}
  h(x,y)\equiv \frac{\pd{\mathcal{H}}{p}}{\bm{l}^{T}F\mathbf{r}_{0}} = \frac{\pd{\mathcal{H}}{q}}{\bm{l}^{T}G\mathbf{r}_{0}}.
\end{equation}
Then, after combining \eqref{eq:50} and \eqref{eq:51} we have that
\begin{equation}
  \label{eq:53}
  \dot{\prefac} = h(x,y)\bm{l}^{T}\left[\left( \pd{F}{x} + \pd{G}{y}\right)\mathbf{r}_{0} + F\pd{\mathbf{r}_{0}}{x} + G \pd{\mathbf{r}_{0}}{y}\right].
\end{equation}
The above requires values of $\pd{\mathbf{r}_{0}}{x}$ and $\pd{\mathbf{r}_{0}}{x}$, which are not provided by the system \eqref{eq:38}.  To obtain these, a formula is needed to relate the Hessian matrix, $Z(x,y)$, of $\Phi(x,y)$ to $\nabla \mathbf{r}_{0}$.  Then, the Hessian matrix can be computed by expanding the system of ray equations \eqref{eq:38}.  

First, differentiate both sides of equation \eqref{eq:33} to get 
\begin{equation}
  \label{eq:54}
  \left[A + pF + qG\right]\nabla \mathbf{r}_{0} = -\left( \nabla A + \nabla (p F) + \nabla (q G) \right) \mathbf{r}_{0}, 
\end{equation}
The Fredholm Alternative Theorem requires that
\begin{equation}
  \label{eq:112}
  \bm{l}^{T}\left( \nabla A + \nabla (p F) + \nabla (q G) \right) \mathbf{r}_{0}=\bm{l}^{T}(\nabla M)\mathbf{r}_{0}=0,
\end{equation}
which is always true since $M\mathbf{r}_{0} = M^{T}\bm{l} = 0$ and
\begin{eqnarray}
  \label{eq:111}
  0&=&\nabla(\bm{l}^{T}M\mathbf{r}_{0}) \\
&=& (\nabla\bm{l}^{T})M\mathbf{r}_{0}  + \bm{l}^{T}(\nabla M)\mathbf{r}_{0} + \bm{l}^{T}M(\nabla\mathbf{r}_{0})\\
&=& \bm{l}^{T}(\nabla M)\mathbf{r}_{0}.
\end{eqnarray}
The general solution to \eqref{eq:54} is 
\begin{equation}
  \label{eq:114}
  \nabla \mathbf{r}_{0} = - M^{\dag}\nabla M \mathbf{r}_{0} +  \alpha \mathbf{r}_{0},
\end{equation}
where $M^{\dag}$ is the pseudoinverse of the matrix $M$ and $\alpha$ is an unknown constant.  Since the vector $\mathbf{r}_{0}$ is normalized so that its entries sum to one, it follows that $\sum_{n} (\nabla\mathbf{r}_{0})_{n} = 0$.  Summing over both sides of equation \eqref{eq:114} then yields
\begin{equation}
  \label{eq:121}
  \alpha = \sum_{n=0}^{2} (M^{\dag}\nabla M \mathbf{r}_{0})_{n}.
\end{equation}
Thus, we have that 
\begin{equation}
  \label{eq:125}
  \nabla \mathbf{r}_{0}  = \mathbf{z}  - \sum_{n=0}^{2}z_{n}\mathbf{r}_{0},\quad \mathbf{z} = - M^{\dag}\nabla M \mathbf{r}_{0}.
\end{equation}

Equation \eqref{eq:125} gives a relationship between $Z$, $\mathbf{r}_{0}$, and $\nabla \mathbf{r}_{0}$.  To obtain the Hessian matrix, $Z(x,y)$, away from the fixed point, the ray equations are extended to include the variables
\begin{eqnarray}
  \label{eq:55}
  x_{j}=\pd{x}{u_{j}},\quad y_{j}=\pd{y}{u_{j}},\\
   p_{j}=\pd{p}{u_{j}},\quad q_{j}=\pd{q}{u_{j}},
\end{eqnarray}
for $j=1,2$.  A good choice for the current problem is to take $u_{1} = x_{0}(\theta)$ and $u_{2}=y_{0}(\theta)$, where $(x_{0}(\theta),y_{0}(\theta))$ is a point on the initial curve defined by \eqref{eq:42}.  The Hessian matrix is then obtained using
\begin{equation}
  \label{eq:56}
  \left[\begin{array}{c c}
    p_{1} & p_{2}\\
    q_{1} & q_{2}
  \end{array}\right] = Z
  \left[\begin{array}{c c}
     x_{1} & x_{2}\\
     y_{1} & y_{2}
  \end{array}\right].
\end{equation}
As long as the matrix on the RHS is invertible, the matrix $Z$ can be obtained along characteristics and $\prefac$ can be integrated numerically using \eqref{eq:53} and \eqref{eq:54}.  The dynamics for the extended variables \eqref{eq:55} is given by
\begin{equation}
  \label{eq:110}
  \dot{\mathbf{v}}_{j}  = J(x,y,p,q)\mathbf{v}_{j},
\end{equation}
where $\mathbf{v}_{j} \equiv (x_{j},y_{j},p_{j},q_{j})^{T}$ and $J(x,y)$ is the Jaciobian matrix for the system \eqref{eq:38}.  One can choose different variables $u_{j}$ with which to extend the system based on what works in practice, and the only thing one must change are the initial conditions.  For our choice, the initial conditions are
\begin{eqnarray}
  \label{eq:113}
  x_{j}(0) = \delta_{1,j},\quad y_{j}(0)=\delta_{2,j},\\
 p_{j}(0) = Z_{j,1}(x_{*},y_{*}) ,\quad q_{j}(0) = Z_{j,2}(x_{*},y_{*}),
\end{eqnarray}
where the matrix $Z(x_{*},y_{*})$ is the solution to \eqref{eq:44}.

\subsubsection{Stability landscape for the full process}
\label{sec:full}

The above analysis can be repeated to obtain a Hamiltonian system for the full process (with protein noise), but a choice must be made for how the limit $\epsilon\to0$, $\gamma\to0$ is taken.  First consider the equation for the quasi-stationary density, $\qss$, for the full process \eqref{eq:7}
\begin{equation}
  \label{eq:57}
   \left[  \frac{1}{\epsilon}A + \frac{1}{\gamma}T \right]\qss(x,y) = 0.
\end{equation}
Here, the domain is the cone $0<x<y<\infty$.  The quasi-stationary density is assumed to have the form
\begin{equation}
  \label{eq:58}
  \qss(x,y) = \mathbf{r}(x,y)\explr{\frac{1}{\epsilon}\Phi(x,y)},
\end{equation}
where again $\mathbf{r}$ is a $3$-vector and $\Phi$ is a scalar function representing the stability landscape.  Note that we have ignored higher order terms here because we only want the Hamiltonian function for comparison to the QD process.  Substituting \eqref{eq:58} into \eqref{eq:57} does not lead to any meaningful equation for $\Phi$ at leading order unless we make an assumption about how the limit $\gamma\to0$ is taken.  
There are two relevant cases: $\gamma = o(\epsilon)$ and $\gamma=\bigo(\epsilon)$.  In the former case, one recovers the QD process \eqref{eq:33}, and in the latter case, collecting terms of leading order in $\epsilon$, with $\gamma=\hgam\epsilon$, yields
\begin{equation}
  \label{eq:59}
  \left[A + \frac{1}{\hgam}\mathbf{H}\right]\mathbf{r} = 0,
\end{equation}
where   $\mathbf{H} \equiv \diag(H_{0}(x,y,p,q),H_{1}(x,y,p,q),H_{2}(x,y,p,q))$ and
\begin{eqnarray} 
  \label{eq:60}
  H_{0}(x,y,p,q) &\equiv& h(x,y,p,q) +\frac{e^{-\hgam p}-1}{\hgam}, \\
  H_{1}(x,y,p,q) &\equiv& h(x,y,p,q) + \frac{e^{-\hgam p}-1}{\hgam}+\frac{e^{-\hgam q}-1}{\hgam},\\
  H_{2}(x,y,p,q) &\equiv& h(x,y,p,q)+\frac{e^{-\hgam q}-1}{\hgam},\\
h(x,y,p,q) &\equiv& \left(\frac{e^{\hgam p}-1}{\hgam}\right)x+\left(\frac{e^{\hgam q}-1}{\hgam}\right)y.
\end{eqnarray}
Thus, the Hamiltonian for the full process is
\begin{equation}
  \label{eq:61}
  \mathcal{H}(x,y,p,q) \equiv \mbox{det}\left[A(x,y)+\mathbf{H}(x,y,p,q)\right],
\end{equation}
which we refer to as the full Hamiltonian

The differences between the full process and the QD process is nicely illustrated by comparing their associated Hamiltonians.  Notice that the full Hamiltonian \eqref{eq:61} is a transcendental function of $p$ and $q$, whereas the Hamiltonian for the QD process \eqref{eq:36} is a cubic polynomial in $p$ and $q$.  One can view this as a Taylor series expansion of the full Hamiltonian about $\hgam(p,q)=(0,0)$.  For this reason, the QD process, as an approximation for the full process with a small amount of protein noise, is only valid within a neighborhood of a deterministic fixed point.  This is, of course, just a reflection of the fluctuation dissipation theorem \cite{gardiner83a}.

\begin{figure}[htb]
  \centering
  \includegraphics[width=12cm]{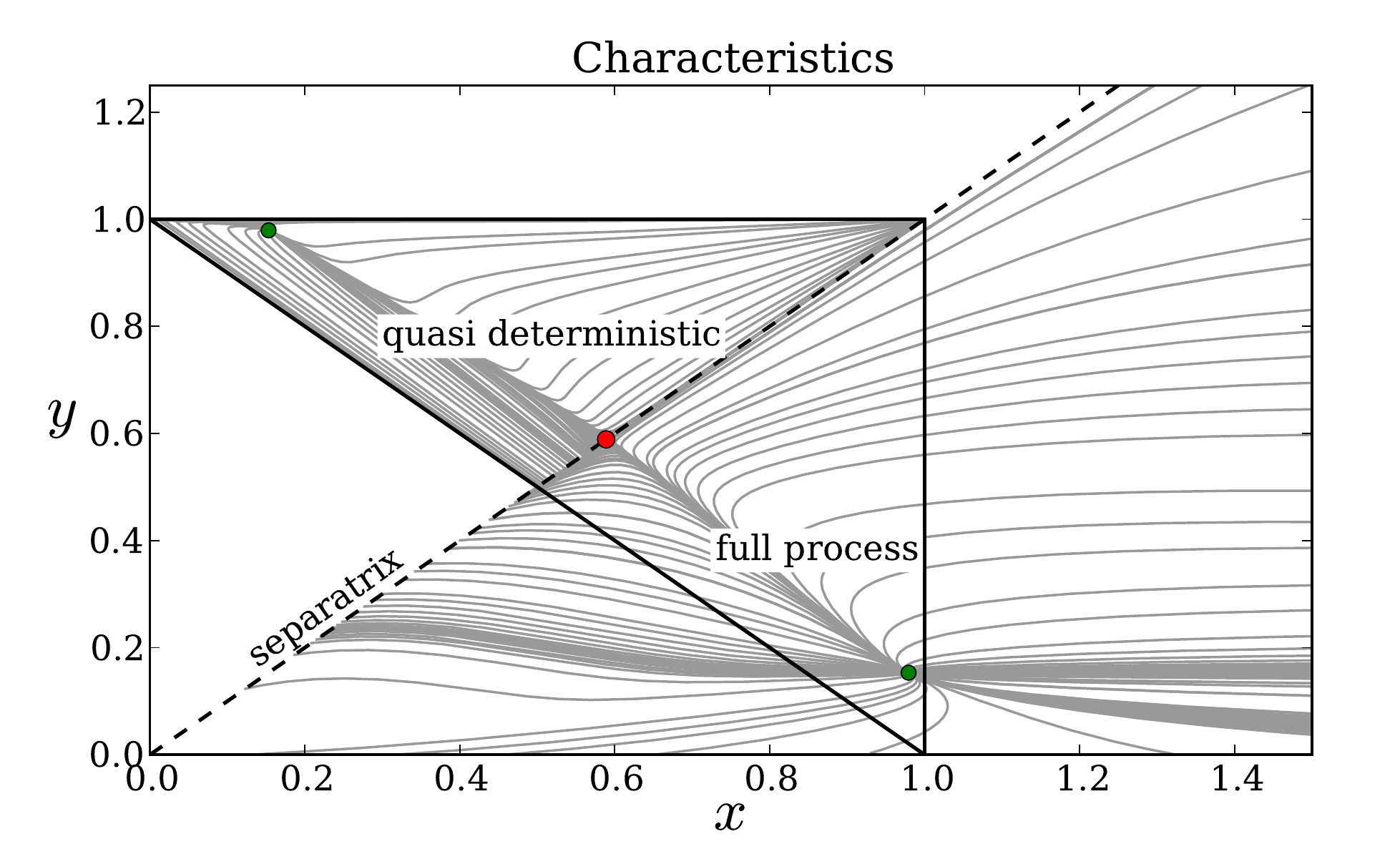}
  \caption{Numerical solutions to the ray equations for the QD and full Hamiltonians.  The grey lines are characteristic projections, $(x(t),y(t))$.  The triangular restricted domain boundary for the QD process is shown with black lines.  The two symmetric stable fixed points are represented by green circles, and the unstable fixed point by a red circle.  The separatrix along which an absorbing boundary condition is imposed is shown by a dashed black line. Parameter values used are $b=1.5$, $\omega = 10^{-4}$ for the QD Hamiltonian, and $\omega=10^{-10}$ for the full Hamiltonian.}
  \label{fig:rays}
\end{figure}

An example of numerical integration (for details regarding numerics see the Appendix) of the ray equations \eqref{eq:38} for the QD \eqref{eq:36} and full \eqref{eq:61} Hamiltonian is shown in Fig.~\ref{fig:rays}.  
The QD rays are shown above the separatrix for comparison.  Notice that the QD rays are contained within a triangular domain, while the rays from the full Hamiltonian cover the entire domain.  This is due to the domain restriction that occurs when removing the protein fluctuations from the process.

\subsection{Adjoint eigenfunction}
Up to terms that are exponentially small in $\epsilon$, the adjoint eigenfunction, $\baefn{0}$, satisfies
\begin{equation}
  \label{eq:66}
  \left[\frac{1}{\epsilon}A^{T} + F\pd{}{x} + G\pd{}{y}\right]\baefn{0} = 0.
\end{equation}
To make things easier, we change coordinates with 
\begin{equation}
  \label{eq:67}
  x = \frac{1}{2}(1+\tau + \sigma),\qquad y = \frac{1}{2}(1+\tau - \sigma),
\end{equation}
so that
\begin{equation}
  \label{eq:68}
  \tau = x+y-1,\qquad \sigma = x-y.
\end{equation}
This transforms the absorbing boundary, $x-y=0$, to the vertical line, $\sigma=0$.  Then, \eqref{eq:66} becomes
\begin{equation}
  \label{eq:69}
    \left[\frac{1}{\epsilon}\hat{A}^{T}(\tau,\sigma) + \hat{F}(\tau)\pd{}{\tau} + \hat{G}(\sigma)\pd{}{\sigma} \right]\baefn{0}(\tau,\sigma) = 0.
\end{equation}
where
\begin{eqnarray}
  \label{eq:70}
\hat{A}(\tau,\sigma) &\equiv& A(x(\tau,\sigma),y(\tau,\sigma)),\\
\hat{F}(\tau) &\equiv&  F+G = \diag(-\tau,1-\tau,-\tau),\\
\hat{G}(\sigma) &\equiv& F-G = \diag(1-\sigma,-\sigma,-(1+\sigma)),
\end{eqnarray}
where $A$, $F$, and $G$ are defined in \eqref{eq:14} and \eqref{eq:15}.  The absorbing boundary condition is then
\begin{equation}
  \label{eq:71}
  \aefn{0}{2}(\tau,0)=0,\quad \mbox{for } \tau\in (-1,1).
\end{equation}

Before proceeding, it is convenient to make the following definitions.  In the rest of this section, we make frequent use of the eigenvectors (right eigenvectors $\bevec{n}$ and left eigenvectors $\blevec{n}$) and eigenvalues, $\mu_{n}$,  satisfying 
\begin{eqnarray}
  \label{eq:72}
  \hat{A}(\tau,\sigma)\bevec{n}(\tau,\sigma) &=& \mu_{n}(\tau,\sigma)\hat{G}(\sigma)\bevec{n}(\tau,\sigma) \\
  \hat{A}(\tau,\sigma)^{T}\blevec{n}(\tau,\sigma) &=& \mu_{n}(\tau,\sigma)\hat{G}(\sigma)\blevec{n}(\tau,\sigma).
\end{eqnarray}
We normalize the two sets of eigenvectors (which are biorthogonal) so that $\blevec{i}^{T}\hat{G}\bevec{j} = \delta_{i,j}$.  Note that because the matrices $\hat{A}$ and $\hat{G}$ are functions of $(\tau,\sigma)$, so are the eigenpairs.  It is easily shown that one of the eigenvalues is zero for all values of $(\tau,\sigma)$, which we set to $\mu_{0}=0$.  The right eigenvector, $\bevec{0}$, is then given by the nullspace of the matrix $\hat{A}$
\begin{equation}
  \label{eq:73}
  \bevec{0}(\tau,\sigma) = \frac{1}{2(\sigma^{2}+(\tau+1)^{2}+2b)}
  \left(\begin{array}{c}
    (1+\tau+\sigma)^{2}\\
    4b\\
    (1+\tau-\sigma)^{2})^{T}
  \end{array}\right)
\end{equation}
Furthermore, the corresponding left eigenvector is simply
\begin{equation}
  \label{eq:74}
  \blevec{0} = \bfone \equiv (1,1,1)^{T}.
\end{equation}
It is convenient to define distinct notation for the eigenpairs evaluated on $\Gamma$, with 
\begin{equation}
  \label{eq:75}
  \bevecg{n}(\tau) \equiv \bevec{n}(\tau,0), \quad \blevecg{n}(\tau) \equiv \blevec{n}(\tau,0), \quad \tilde{\mu}_{n}(\tau)\equiv\mu_{n}(\tau,0).
\end{equation}
At the boundary one of the eigenvalues, $\mu_{1}$ say, vanishes and the eigenspace for the zero eigenvalue is degenerate (i.e. there are two zero eigenvalues but the nullspace is one dimensional) which means that $\tilde{\mu}_{1} = 0$, $\bevecg{1}=\bevecg{0}$ and $\blevecg{1}=\bfone$.

The approximation of the adjoint eigenfunction proceeds using singular perturbation methods, along the lines of \cite{keener11a}.  Three solutions are found which are valid in different regions of the domain: an outer solution, a boundary layer solution for the $\bigo(\epsilon)$ strip near the absorbing boundary, and a transition layer solution in the $\bigo(\epsilon^{1/2})$ overlap region between the other two.

Away from the boundary, the exact solution (that does not satisfy the boundary condition) is
\begin{equation}
  \label{eq:76}
  \baefn{\mathrm{out}} = \bfone.
\end{equation}
To obtain a uniform asymptotic approximation that also satisfies \eqref{eq:71}, a boundary-layer solution is needed.  Consider the stretched variable $z = \sigma/\epsilon$.  To leading order in $\epsilon$, the boundary-layer solutions, $\baefn{\mathrm{bl}}(\tau,z)$, satisfies
\begin{equation}
  \label{eq:77}
     \hat{G}(0)\pd{\baefn{\mathrm{bl}}}{z} + \hat{A}(\tau,0)^{T}\baefn{\mathrm{bl}} = 0,
\end{equation}
where $\hat{G}(0) = \diag(1,0,-1)$.  The solution has the form
\begin{equation}
  \label{eq:78}
  \baefn{\mathrm{bl}}(\tau,z) = c_{0}\bfone + c_{1}(\bm{\zeta} + z\bfone) + c_{2}\blevecg{2} e^{-\tilde{\mu}_{2}z},
\end{equation}
where $\blevecg{2}$, $\tilde{\mu}_{2}$ is the only eigenpair (on the boundary) with a nonzero eigenvalue.  However, the eigenvalue, $\tilde{\mu}_{2}(\tau)$, is negative for all values of $\tau\in(-1,1)$, and in order to obtain a bounded solution in the limit $z\to\infty$ we set $c_{2}=0$.  The vector $\bm{\zeta}$ is the generalized left eigenvector satisfying $\hat{A}(\tau,0)^{T}\bm{\zeta} = \hat{G}(0)\bfone$ and is given by
\begin{equation}
  \label{eq:79}
  \bm{\zeta} = (-1/b,0,1/b)^{T}.
\end{equation}
At the boundary, the solution is
\begin{equation}
  \label{eq:80}
  \baefn{\mathrm{bl}}(\tau,0) = c_{0}\bfone + c_{1}\bm{\zeta},
\end{equation}
and the boundary condition \eqref{eq:71} requires
\begin{equation}
  \label{eq:81}
  c_{0} + \frac{c_{1}}{b} = 0,
\end{equation}
so that $ c_{1} = -bc_{0}$.  Thus, up to a single unknown constant, $c_{0}$, which must be determined by matching, the boundary-layer solution is
\begin{equation}
  \label{eq:82}
  \baefn{\mathrm{bl}}(\tau,z) = c_{0}(\bfone - b(\bm{\zeta} + z\bfone)).
\end{equation}
Because $\baefn{\mathrm{bl}}(\tau,z)$ is unbounded in the limit $z\to \infty$, it is not possible to match it to the outer solution, $\baefn{\mathrm{out}}$.  We can think of the term, $\bm{\zeta}+z\bfone$, in the boundary-layer solution is a truncated Taylor series expansion of the true solution around $z=0$.  To match the boundary-layer and outer solutions, a transition-layer solution is required for the strip of width $\bigo(\sqrt{\epsilon})$ along the boundary.  

Consider the stretched coordinate $s = \sigma/\sqrt{\epsilon}$.  Keeping terms to $\bigo(\epsilon^{1/2})$, the transition-layer solution, $\baefn{\mathrm{tl}}(\tau,s)$, satisfies
\begin{equation}
  \label{eq:83}
  \sqrt{\epsilon}\hat{G}(0)\pd{\baefn{\mathrm{tl}}}{s} + \hat{A}(\tau,0)^{T}\baefn{\mathrm{tl}} = 0.
\end{equation}
It is less clear how to truncate the above equation to obtain a leading order transition-layer solution.  Because we must match the outer solution, $\bfone$, to the boundary layer solution that has the generalized eigenvector $\bm{\zeta}$, we try a solution of the form
\begin{equation}
  \label{eq:84}
  \baefn{\mathrm{tl}}(\tau,s) = a_{0}(\tau,s)\bfone + a_{1}(\tau,s)\bm{\chi}(\tau,s),
\end{equation}
where 
\begin{equation}
  \label{eq:85}
  \bm{\chi}(\tau,s)\equiv \frac{1}{\mv(\tau,\sqrt{\epsilon}s)}(\bfone - \blevec{1}(\tau,\sqrt{\epsilon}s)),
\end{equation}
and $a_{0,1}$ are unknown scalar functions.   In the limit $\sigma\to0$, the deterministic flux across the boundary,
\begin{equation}
  \label{eq:86}
  \mv(\tau,\sigma)\equiv  f(x(\tau,\sigma),y(\tau,\sigma)) - f(y(\tau,\sigma),x(\tau,\sigma)),
\end{equation}
(with $f(x,y)$ given by \eqref{eq:12})
vanishes and the eigenvector $\blevec{1} \to \bfone$. That is, the eigenvalue $\mu_{1}$, corresponding to the eigenvector $\blevec{1}$, vanishes on the boundary.  Furthermore, it can be shown that
\begin{equation}
  \label{eq:115}
  \lim_{s\to0}\bm{\chi}(\tau,s) = -\frac{\tilde{\mu}_{1}^{(\sigma)}(\tau) }{\tilde{\mv}^{(\sigma)}(\tau)}\bm{\zeta},
\end{equation}
where
\begin{equation}
  \label{eq:92}
  \tilde{\mv}^{(\sigma)}(\tau) \equiv \pd{}{\sigma}\mv(\tau,0),\quad \tilde{\mu}_{1}^{(\sigma)}(\tau) \equiv \pd{}{\sigma}\mu_{1}(\tau,0).
\end{equation}
Substituting \eqref{eq:84} into \eqref{eq:83} yields
\begin{equation}
  \label{eq:87}
 \sqrt{\epsilon} \hat{G}(0)\left( \pd{a_{0}}{s} \bfone + \pd{a_{1}}{s}\bm{\chi} + \sqrt{\epsilon}a_{1}\pd{\bm{\chi}}{\sigma}  \right) + a_{1}\hat{A}(\tau,0)^{T}\bm{\chi} = 0.
\end{equation}
To obtain the unknown functions $a_{0,1}(\tau,s)$ we project \eqref{eq:87} with the right eigenvectors, $\bevec{n}(\tau,s)$, $n=0,1$.  After applying these projections (using the fact that $\bevec{0}^{T}\hat{G}\bm{\chi} = 1$, $\bevec{0}^{T}\hat{A}^{T} = 0$, $\bevec{0}^{T}\hat{G}\bfone = \mv(\tau,\sigma)$, and $\bevec{1}^{T}\hat{G}\bfone = 0$) and collecting leading order terms in $\epsilon$, we get
\begin{eqnarray}
  \label{eq:88}
  \pd{a_{1}}{s} = \mv(\tau,\sqrt{\epsilon}s) \pd{a_{0}}{s}\\
  \label{eq:89}
  \pd{a_{1}}{s} + s\tilde{\mu}_{1}^{(\sigma)}(\tau) a_{1} = 0,
\end{eqnarray}
where
\begin{equation}
  \label{eq:90}
  \tilde{\mu}_{1}^{(\sigma)}(\tau)  = -\frac{b(\tau^{2}+2b-1)}{(\tau+1)^{2}},
\end{equation}
 It turns out that $\tilde{\mu}_{1}^{(\sigma)}$ is related to the curvature of the stability landscape normal to the separatrix; that is, if we define
\begin{equation}
  \label{eq:117}
  \hat{\Phi}(\tau,\sigma) \equiv \Phi(x(\tau,\sigma),y(\tau,\sigma))
\end{equation}
then
\begin{equation}
  \label{eq:116}
  \tilde{\mu}_{1}^{(\sigma)}(\tau) = - \pdd{}{\sigma}\hat{\Phi}(\tau,0).
\end{equation}
At $\tau_{\alpha} = \sqrt{1-2b}$, the curvature vanishes and changes sign but is always negative (with $\tilde{\mu}_{1}'(\tau_{u})>0$) at the unstable fixed point, $(\tau_{u},0)$.  Divide the separatrix ($\sigma=0$ and $-1<\tau<1$) into three regions: $-1<\tau<0$, $0<\tau<\tau_{\alpha}$, and $\tau_{\alpha}<\tau<1$.  The first region is ignored because it is in part of the domain $\mathcal{D}$ excluded from the stationary density function (see Sec.~\ref{sec:qsd}).  The second region contains the unstable fixed point, and the third we can ignore as only extremely rare trajectories cross the separatrix in this region.
Up to an unknown constant, the solutions to \eqref{eq:88} and \eqref{eq:89} are
\begin{eqnarray}
  \label{eq:91}
 a_{0}(\tau,s) &\sim& -\hat{a}\frac{\tilde{\mu}_{1}^{(\sigma)}(\tau) }{\epsilon^{1/2}\tilde{\mv}^{(\sigma)}(\tau)}\int_{0}^{s}\explr{-\frac{1}{2}\tilde{\mu}_{1}^{(\sigma)}(\tau) u^{2}}du,\\
  a_{1}(\tau,s) &\sim& \hat{a}\explr{-\frac{1}{2}\tilde{\mu}_{1}^{(\sigma)}(\tau) s^{2}}.
\end{eqnarray}
The transition layer solution is then
\begin{eqnarray}
  \label{eq:93}
  \baefn{\mathrm{tl}}(\tau,s) &=& \hat{a}\left(-\frac{\tilde{\mu}_{1}^{(\sigma)}(\tau) }{\epsilon^{1/2}\tilde{\mv}^{(\sigma)}(\tau)}\int_{0}^{s}\explr{-\frac{1}{2}\tilde{\mu}_{1}^{(\sigma)}(\tau) u^{2}}du\bfone \right. \\
\nonumber
   &&\qquad\left. + \explr{-\frac{1}{2}\tilde{\mu}_{1}^{(\sigma)}(\tau) s^{2}}\bm{\chi}(\tau,s)\right).
\end{eqnarray}

The three solutions can now be matched.  
First, matching the transition layer solution to the boundary layer solution is done using the Van-Dyke rule.  In terms of the boundary layer variable, $z$, the transition layer solution is
\begin{equation}
  \label{eq:94}
  \baefn{\mathrm{tl}}(\tau,\epsilon^{1/2}z) \sim -\hat{a}\frac{\tilde{\mu}_{1}^{(\sigma)}(\tau) }{\tilde{\mv}^{(\sigma)}(\tau)} (\bm{\zeta} + z\bfone),
\end{equation}
Matching terms with the boundary layer solution yields
\begin{equation}
  \label{eq:95}
  \hat{a} = bc_{0}\frac{\tilde{\mv}^{(\sigma)}(\tau)}{\tilde{\mu}_{1}^{(\sigma)}(\tau) }.
\end{equation}
The composite boundary/transition layer solution is then
\begin{eqnarray}
  \label{eq:96}
    \baefn{\mathrm{bl/tl}}(\tau,\sigma) &=& c_{0}\left[\bfone - b\left(\epsilon^{-1/2}\int_{0}^{\sigma/\sqrt{\epsilon}}\explr{-\frac{1}{2}\tilde{\mu}_{1}^{(\sigma)}(\tau) u^{2}}du\bfone\right.\right.\\
\nonumber
     &&\qquad\qquad \left.\left.+\explr{-\frac{1}{2}\tilde{\mu}_{1}^{(\sigma)}(\tau) \frac{\sigma^{2}}{\epsilon}}\frac{\tilde{\mv}^{(\sigma)}(\tau)}{\tilde{\mu}_{1}^{(\sigma)}(\tau) }
                                        \bm{\chi}(\tau,\frac{\sigma}{\sqrt{\epsilon}})\right)\right]
\end{eqnarray}

The final unknown constant, $c_{0}$, is determined by matching to the outer solution so that
\begin{equation}
  \label{eq:97}
  \lim_{s\to \infty}\baefn{\mathrm{bl/tl}}(\tau,s) = \bfone,
\end{equation}
which implies that
\begin{equation}
  \label{eq:98}
  c_{0} =  -\frac{\sqrt{2\epsilon \tilde{\mu}_{1}^{(\sigma)}(\tau) }}{b \sqrt{\pi} - \sqrt{2\epsilon \tilde{\mu}_{1}^{(\sigma)}(\tau) }}.
\end{equation}
In order to evaluate the term in the numerator of the eigenvalue formula \eqref{eq:30}, we require the adjoint eigenfunction evaluated on the boundary (in a neighborhood of the unstable fixed point) which is
\begin{equation}
  \label{eq:99}
  \baefn{0}(\tau,0) \sim \frac{\sqrt{2\epsilon \tilde{\mu}_{1}^{(\sigma)}(\tau) }}{b \sqrt{\pi} - \sqrt{2\epsilon \tilde{\mu}_{1}^{(\sigma)}(\tau) }}\left(-\bfone + b\bm{\zeta}\right), \quad \tau \in (0,\sqrt{1-2b}).
\end{equation}

\subsection{Principal eigenvalue}
\label{sec:eval}
We now have all of the components necessary to approximate the principal eigenvalue, using the formula \eqref{eq:30}.  First, for the term in the denominator, we can approximate the adjoint eigenfunction with the outer solution $\baefn{0}\sim \bfone$ and the (unnormalized) stationary density with \eqref{eq:32} (the higher order term $\mathbf{r}_{1}$ can be ignored).  Then the term in the denominator is simply the normalization factor, which can be approximated using Laplace's Method to get
\begin{equation}
  \label{eq:100}
  \int_{\mathcal{D}} \explr{-\frac{1}{\epsilon}\Phi(x,y) - \prefac(x,y)} dA \sim \frac{2\pi\epsilon}{\sqrt{\mbox{det}(Z)}},
\end{equation}
where $Z$ is the Hessian matrix \eqref{eq:40} of $\Phi$ at the stable fixed point $(x_{*},y_{*})$.  Note that we have used the fact that $\prefac(x_{*},y_{*})=\Phi(x_{*},y_{*})=0$ and that the vector $\mathbf{r}_{0}$ is normalized so that its entries sum to one.  

The term in the numerator of \eqref{eq:32} requires the approximation \eqref{eq:99} of the adjoint eigenfunction on the absorbing boundary.  The integral can also be approximated using Laplace's Method, with
\begin{eqnarray}
  \label{eq:101}
  && \oint_{\Gamma}\baefn{0}^{T}(F-G)\qss\; ds\\
\nonumber
    &\sim&\frac{1}{\sqrt{2}}\int_{0}^{1}\frac{b\sqrt{2\epsilon \tilde{\mu}_{1}^{(\sigma)}(\tau) }}{b \sqrt{\pi} - \sqrt{2\epsilon\tilde{\mu}_{1}^{(\sigma)}(\tau) }} \bm{\zeta}^{T}\hat{G}(0)\tilde{\mathbf{r}}_{0}(\tau) \explr{-\frac{1}{\epsilon}\tilde{\Phi}(\tau) - \tilde{\prefac}(\tau)}d\tau \\
\nonumber
& \sim& \frac{\epsilon b\sqrt{\pi}e^{-\prefac(x_{u},y_{u})}}{b\sqrt{\pi} - \epsilon \sqrt{2\tilde{\mu}_{1}^{(\sigma)}(\tau_{u})}} \sqrt{\frac{2\tilde{\mu}_{1}^{(\sigma)}(\tau_{u})}{\tilde{\Phi}''(\tau_{u})}}\bm{\zeta}^{T}\hat{G}(0)\tilde{\mathbf{r}}_{0}(\tau)\explr{-\frac{1}{\epsilon}\Phi(x_{u},y_{u})}.
\end{eqnarray}
For convenience, we have defined functions on the boundary in the variable $\tau$ with
\begin{eqnarray}
  \label{eq:102}
  \tilde{\mathbf{r}}_{0}(\tau) &\equiv& \mathbf{r}_{0}(x(\tau,0),y(\tau,0)) \\
  \tilde{\Phi}(\tau) &\equiv& \Phi(x(\tau,0),y(\tau,0)) \\
  \tilde{\prefac}(\tau) & \equiv& \prefac(x(\tau,0),y(\tau,0)),
\end{eqnarray}
where $x(\tau,\sigma)$ and $y(\tau,\sigma)$ are defined in \eqref{eq:67}.  

Although the quantities $\Phi(x_{u},y_{u})$ and $\prefac(x_{u},y_{u})$ must be computed numerically, the remaining unknown terms can be computed analytically by exploiting the reflection symmetry of the problem.  Along $\Gamma$, we have that $x=y$ and $p=q$ so that the equation for $\Phi$ and $\mathbf{r}_{0}$ \eqref{eq:33} can be written as
\begin{equation}
  \label{eq:103}
  \left[\hat{A}(\tau,0) - \mu_{\Gamma}(\tau) \hat{F}(\tau) \right]\tilde{\mathbf{r}}_{0}(\tau) = 0,
\end{equation}
where we have defined
\begin{equation}
  \label{eq:127}
  \mu_{\Gamma}(\tau) \equiv - p(x(\tau,0),y(\tau,0))=-q(x(\tau,0),y(\tau,0)).
\end{equation}
The stability landscape function on the boundary is then
\begin{equation}
  \label{eq:104}
  \tilde{\Phi}(\tau) = \Phi(x_{u},y_{u}) - \int_{\tau_{u}}^{\tau}\mu_{\Gamma}(u)du.
\end{equation}
The above is just an eigenvalue problem with three possible solutions, one of which can be excluded because there is a zero eigenvalue corresponding to the nullspace of $\hat{A}$.  It can be shown \cite{newby11b} that if the diagonal elements of $\hat{F}(\tau)$ are such that at least two have oposite sign then only one of the remaining two eigenvalues has a corresponding positive eigenvector ($\tilde{\mathbf{r}}_{0}(\tau)$ must have positive elements) making the solution to \eqref{eq:103} unique.  It turns out that this is only true for $\tau\in(0,1)$, which is due to the domain restriction caused by removing protein fluctuations.  The result is
\begin{eqnarray}
  \label{eq:105}
  \mu_{\Gamma}(\tau) &=& - \frac{\tau^{3} + 2\tau^{2} + 2b(\tau-1) + \tau}{2\tau(1-\tau)},\\
\label{eq:106}
  \tilde{\mathbf{r}}_{0}(\tau) &=& \left( \frac{1}{2}(1-\tau),\tau,\frac{1}{2}(1-\tau)  \right)^{T}.
\end{eqnarray}
We also have that
\begin{equation}
  \label{eq:107}
  \tilde{\Phi}''(\tau) = -\mu_{\Gamma}'(\tau) = \frac{-\tau^{4} + 2\tau^{3} + (3+2b)\tau^{2}-4b\tau + 2b}{2\tau^{2}(1-\tau)^{2}}.
\end{equation}
And finally, using \eqref{eq:79}, \eqref{eq:70}, and \eqref{eq:106} we get
\begin{equation}
  \label{eq:108}
  \bm{\zeta}^{T}\hat{G}(0)\tilde{\mathbf{r}}_{0}(\tau) = \frac{1}{b}(1-\tau).
\end{equation}
Combining these components together, we have the final result that
\begin{equation}
  \label{eq:109}
  \fl \lambda_{0}\sim\frac{1}{\sqrt{2\pi}}\left[\frac{ (1-\tau_{u})e^{-\prefac(x_{u},y_{u})}}{b\sqrt{\pi} - \epsilon \sqrt{2\tilde{\mu}_{1}^{(\sigma)}(\tau_{u})}}\right] \sqrt{\frac{\tilde{\mu}_{1}^{(\sigma)}(\tau_{u})\mbox{det}(Z)}{\tilde{\Phi}''(\tau_{u})}} 
           \explr{-\frac{1}{\epsilon}\Phi(x_{u},y_{u})},
\end{equation}
where $\tau_{u} = x_{u}+y_{u}-1$, and $\tilde{\mu}_{1}^{(\sigma)}(\tau) $ is given by \eqref{eq:90}.  As expected, the eigenvalue is exponentially small in $\epsilon$, which means that the height of the stability lanscape at the unstable fixed point, $\Phi(x_{u},y_{u})$, must be approximated as accurately as possible.  

The remaining terms are often referred to as the `prefactor', and except for the quantity $\prefac(x_{u},y_{u})$, all of the terms in the prefactor (the derivatives of the stability landscape function and \eqref{eq:108}) represent properties local to the fixed points.   The remaining term in the prefactor depends on the function $\prefac(x,y)$, which depends on properties of the process not local to the fixed points and must be computed numerically.

\section{Results}
\label{sec:results}
\begin{figure}[htbp]
  \centering
  \includegraphics[width=12cm]{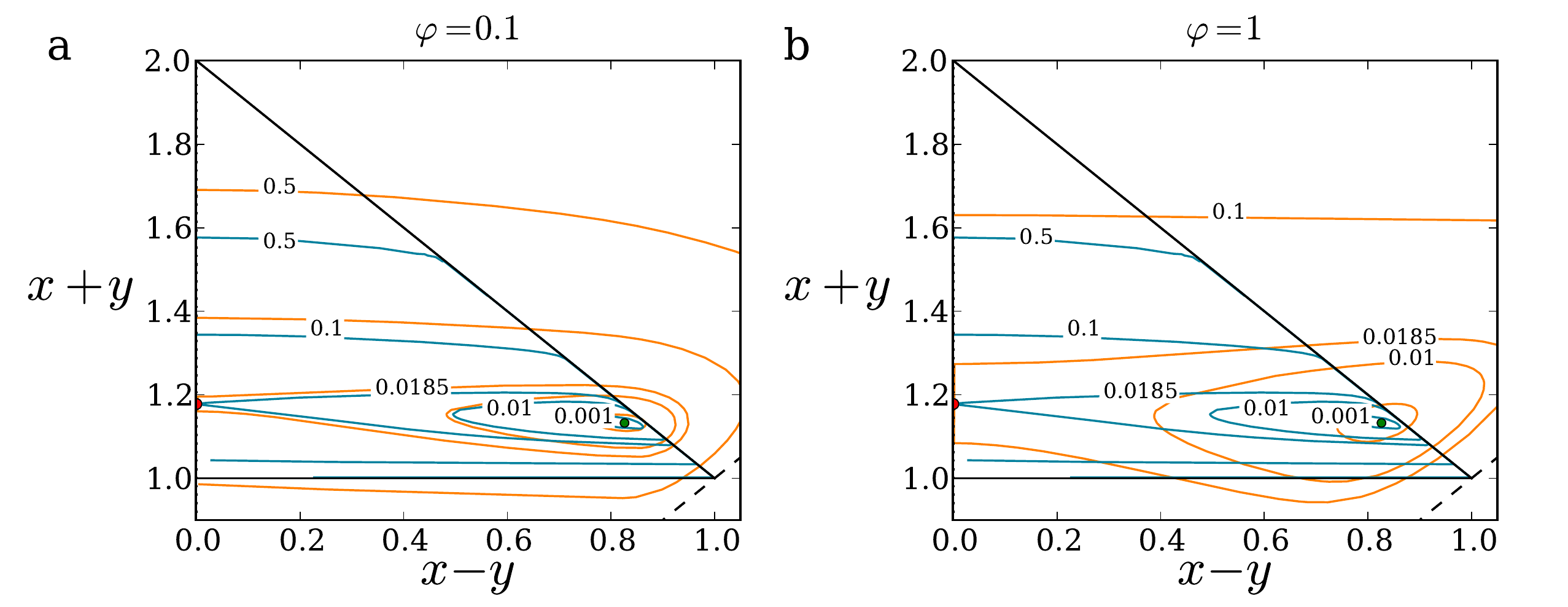}
  \caption{Level curves of the QD (blue) and full process (orange) stability landscape function $\Phi$. Black lines show the domain boundary for the QD process, and the dashed line shows the boundary of positive protein levels.  (a) The QD result is compared to the full process with $\hgam =0.1$ so that the protein noise is weak compared to promotor noise.  (b) Same as (a) but with $\hgam=1$ so that the protein noise strength is comparable to promotor noise.  Parameter values are the same as Fig.~\ref{fig:rays}.}
  \label{fig:level_curves}
\end{figure}

In this section, the results gathered throughout this paper are used to explore how removing the intrinsic noise that arises from protein production/degradation effects the random process.  In particular, we examine the stability landscape and the metastable transition times.  First, in Fig.~\ref{fig:level_curves} the numerical solutions to the ray equations \eqref{eq:38} are used to generate level curves of the stability landscape function, $\Phi$, for both the QD \eqref{eq:36} and the full \eqref{eq:61} Hamiltonians.  
For presentation, the level curves are shown in the $(x+y,x-y)$ plane, with the separatrix along the left edge ($x-y=0$) of each frame.  In Fig.~\ref{fig:level_curves}a, level curves for the full process are shown for $\hgam=0.1$ so that the protein noise is small compared to promotor noise.  Recall that the parameter $\hgam$ controls the strength of protein noise relative to the strength of promotor noise so that there is no protein noise in the limit $\hgam\to0$.  The resulting curves match closely in a neighborhood of the stable fixed point and extend out toward the unstable saddle point.  As expected, the level curves begin to diverge the farther away from either fixed point they are.   In Fig.~\ref{fig:level_curves}b, the strength of the protein noise is increased, with $\hgam = 1$, and in this case, the stability landscape of the two processes are quite different, matching only near the fixed points of the deterministic dynamics.  

As a result of the domain restriction effect, level curves of the full process extend into regions the QD process is excluded from.  This is because protein levels can only cross above the line $y=1-x$, not below it, when there is no protein noise.  This is also true of the lines $x=1$ and $y=1$.
The domain restriction effect is eliminated when protein noise is added back into the process, even if it is very small compared to promotor noise.   For the model considered here, the effect is of no serious consequence for metastable transitions as the restricted domain still contains all three fixed points.  However, a model of a more complex gene circuit might be significantly affected by removing protein noise---especially if this restricts the domain for the protein levels in such a way as to generate qualitatively different behavior, which would imply a nontrivial contribution of protein noise, no matter how negligible it may be.

Because of the symmetry in the problem, we obtained analytical results for various quantities on the separatrix, including the shape stability landscape.  We can use these results to obtain an analytical approximation of the probability density for the position along the separatrix a trajectory passes through as it transitions from one basin of attraction to another.  Using the results of Sec.~\ref{sec:eval}, the stationary density along the separatrix is given by
\begin{equation}
  \label{eq:123}
  \mathbf{p}_{\mathrm{exit}}(\tau) \sim \frac{\tilde{\mathbf{r}}_{0}(\tau)e^{-\tilde{\prefac}(\tau)} \explr{-\frac{1}{\epsilon}\tilde{\Phi}(\tau)}}{\int_{0}^{1}e^{-\tilde{\prefac}(\tau)}\explr{-\frac{1}{\epsilon}\tilde{\Phi}(\tau)}d\tau},
\end{equation}
where we remind the reader that $\tau = x+y-1$ and $x=y$ along the separatrix.  The only term that cannot be obtained analytically is the function $\tilde{\prefac}(\tau)$, which can be ignored as a first approximation.  For simplicity, we also average over the promotor state to get the scalar marginal probability density for the exit point.  Then, using Laplace's method, the exit density is
\begin{equation}
  \label{eq:124}
  P_{\mathrm{exit}}(\tau) \sim \sqrt{\frac{\tilde{\Phi}''(\tau_{u})}{2\pi \epsilon}}\explr{-\frac{1}{\epsilon}(\tilde{\Phi}(\tau)-\tilde{\Phi}(\tau_{u}))},\quad \tau\in(0,1),
\end{equation}
where $\tilde{\Phi}(\tau)$ is given by 
\begin{equation}
  \label{eq:52}
  \tilde{\Phi}(\tau) = -\frac{\tau^{2}}{4} - \frac{3\tau}{2} - b\log(\tau) - 2\log(1-\tau).
\end{equation}

\begin{figure}[tbh]
  \centering
  \includegraphics[width=10cm]{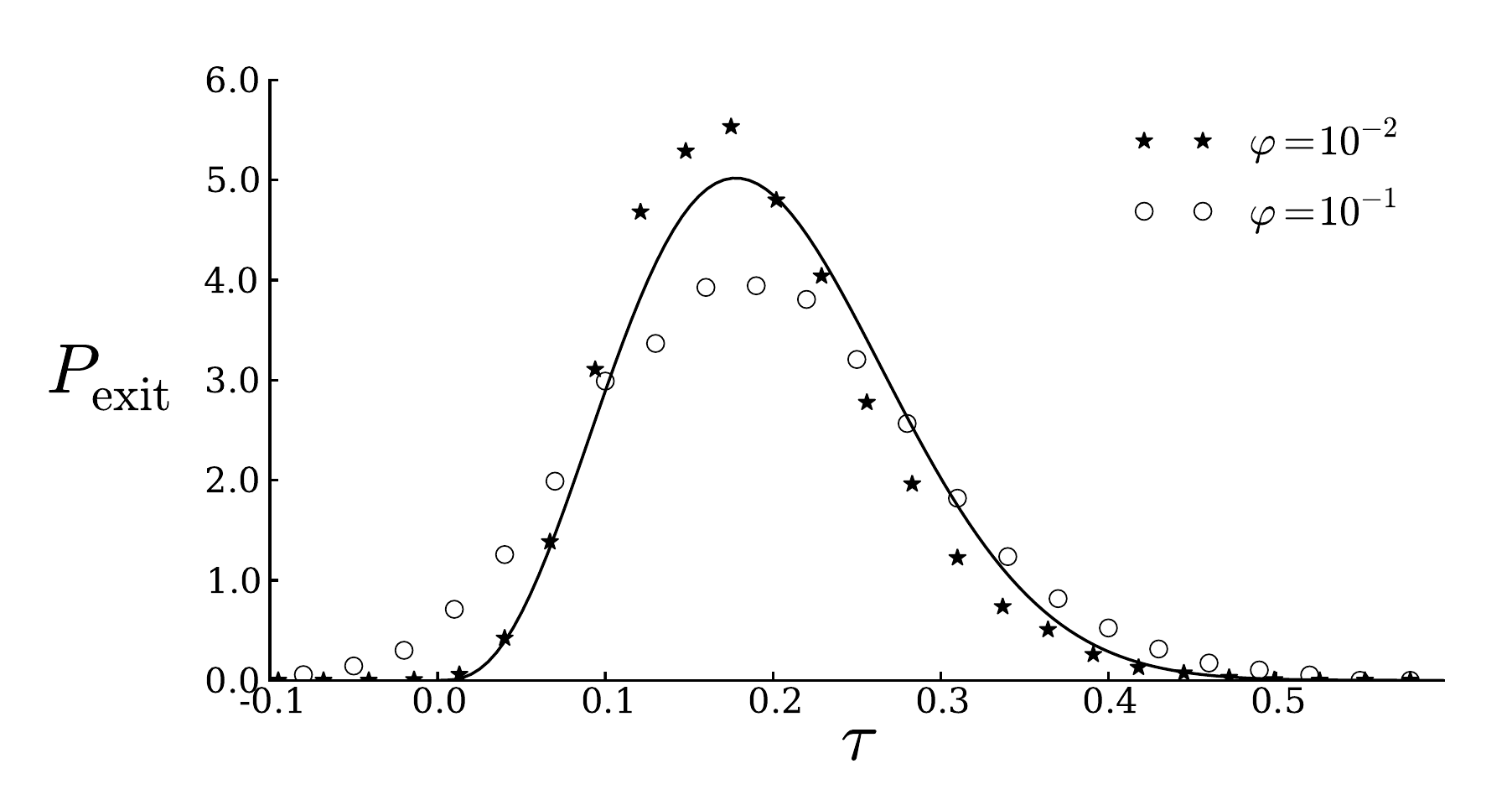}
  \caption{The density of exit points along the separatrix as a function of $\tau = x+y-1$, with $x=y$.  The black curve shows the analytical approximation for the QD ($\hgam=0$) process, and the symbols show histograms from $10^{4}$ Monte-Carlo simulations for different values of $\hgam$.}
  \label{fig:exitdist}
\end{figure}

The QD exit density approximation is shown in Fig.~\ref{fig:exitdist} along with two histograms obtained from Monte-Carlo simulations.  
While the histogram for $\hgam=0.1$ is close to the QD approximation, it is evident that some trajectories pass through the separatrix in the interval $(-1,0)$, which is impossible without protein noise.  This effect becomes negligible when the protein noise is reduced to $\hgam=0.01$.

\begin{figure}[tbh]
  \centering
  \includegraphics[width=12cm]{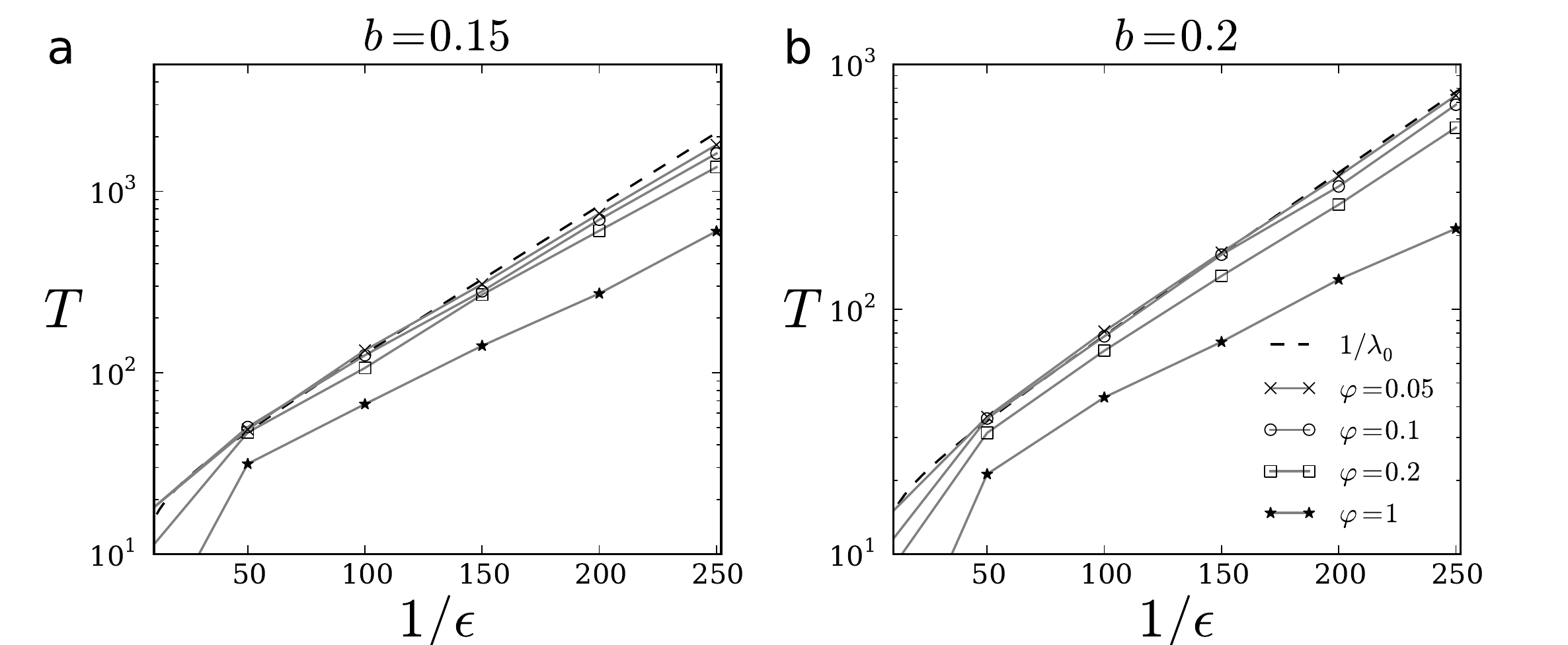}
  \caption{The mean exit time on a log scale as a function of $1/\epsilon$.  Symbols represent $10^{3}$ averaged Monte Carlo simulations for various values of $\hgam$.  The dashed black line is the approximation, $T=1/\lambda_{0}$, for the QD process ($\hgam=0$).}
  \label{fig:mfpt}
\end{figure}

The FETD for a trajectory, starting from a stable fixed point, to reach the separatrix is asymptotically exponential in the large time limit, and the timescale is determined by the principal eigenvalue, $\lambda_{0}$, from equation \eqref{eq:109}.  We can then approximate the mean exit time with $T\sim 1/\lambda_{0}$.  In Fig.~\ref{fig:mfpt} the mean exit time is shown on a log scale as a function of $1/\epsilon$ along with results from Monte-Carlo simulations. 
Notice that the approximation and the Monte-Carlo simulations are asymptotically linear as $1/\epsilon\to\infty$.  For the approximation, the slope of this line is determined by the height of the stability barrier, $\Phi(x_{u},y_{u})$, while the prefactor affects the vertical shift.  From the Monte Carlo results (symbols with grey lines) we see that the mean exit time converges to the approximation as $\hgam\to 0$.  However, it is clear that the slope of the analytical curve is slightly different then that of the Monte Carlo results even when $\hgam$ is small.  Thus, we may think of the mean exit time approximation for the QD process as an asymptotic approximation of the full process in terms of the small parameter $\hgam\ll 1$ so long as $\epsilon$ is also small but not too small.

\section{Discussion}
Understanding how different noise sources affect the dynamics of a gene circuit is essential to understand how different regulatory components interact to produce the complex variety of environmental responses and behaviors.  Even if one excludes extrinsic noise sources---such as environmental and organism-to-organism variations---there are several sources of intrinsic noise, such as fluctuations in translation, transcription, and the conformational state of DNA regulatory units.  

The behavior we are interested in understanding is a transition from one metastable state to another.  Each metastable state corresponds to the stable steady-state solutions of the underlying deterministic system.  The bistable mutual repressor model has two identical stable steady states separated by an unstable saddle node.  On small time scales, the protein levels fluctuate near one of the two stable steady states.  On large time scales, fluctuations cause a metastable transition to occur, where the protein levels shift to the other steady state by crossing the separatrix containing the unstable saddle point.  

To understand how metastable transitions can be induced by promotor noise, we consider a discrete stochastic model of a mutual repressor circuit where the protein levels change deterministically, which we call the QD process.  This is done by fixing the promotor state and then taking the thermodynamic, or large system size, limit of the protein production/degradation reactions.  We then compare the QD process to the full process that includes protein noise.

We find important qualitative differences that persist even when the magnitude of the protein noise is small compared to promotor noise.  In particular, without protein noise and after initial transients, the protein levels are restricted such that the total amount of protein is never less than half its maximum possible value; that is, the total number of proteins is such that $n+m>\alpha/\delta$, where $\alpha$ is the protein production rate and $\delta$ is the degradation rate.  Said another way, assuming that the random process starts at one of the deterministic stable fixed points, promotor fluctuations could never push the protein copy numbers so that, for example, only a single copy of each protein is present.  In contrast, the stability landscape for the full process and Monte-Carlo simulations show that protein levels are able to reach all positive values.  While this restriction does not stop the QD process from exhibiting metastable transitions, more complex gene circuits may require protein fluctuations, even if they are very small, in order to function correctly.

\section*{Appendix}
\label{sec:appendix}
In this appendix, we summarize the numerical methods and tools used throughout the paper.  Most numerical work is performed in Python, using the Numpy/Scipy package.  For more computationally-expensive tasks, we use Scipy's Weave package to include functions written in C, which allows us to use the GNU Scientific Library for numerical Integration of the ray equations \eqref{eq:38},  and for random number generators used in Monte Carlo simulations.  

There are a few notable observations regarding integration of the ray equations.  First, characteristic projections, $(x(t),y(t))$, have a tendency to 'stick' together along certain trajectories, peeling off one at a time (see Fig.~\ref{fig:rays}).  To adequately cover the domain with rays, a shooting method must be used to select points on the Cauchy data.  For more details on this see Ref.~\cite{ludwig75a}.  We found that the simplest method was to use the secant method (we use the ``brentq'' function in the Scipy.integrate package) to minimize the euclidian distance between the final value of $(x(t),y(t))$ along the separatrix and the saddle node.  This method is convenient since it does not require knowledge of the Hessian matrix, $Z(x,y)$.  Second, the value of $\omega$ used to generate Cauchy data must be chosen small enough to get accurate results.  However, we found that if it is chosen too small, rays are no longer able to cover the domain, and more importantly, we could no longer generate a ray that reaches the unstable fixed point on the separatrix.  

For the mutual repressor model, the trajectory connecting the stable fixed point to the saddle is one of the curves along which characteristics tend to stick to each other.  Suppose that $\theta_{m}$ is the point on the Cauchy data \eqref{eq:126} that generates the ray that connects the fixed points.  Then small perturbations $\theta = \theta_{m}+\delta\theta$, $\abs{\delta\theta}\ll 1$, cause the characteristic, $(x(t),y(t))$, to diverge sharply away from the saddle.  This not only makes it difficult to compute $\theta_{m}$, but also creates difficulties for computing the function $k(x,y)$ (see Sec.~\ref{sec:prefac}).  Since the expanded set of ray equations (\ref{eq:110}) track the derivatives of $x$, $y$, $p$, and $q$ with respect to the point on the Cauchy data, which, for values of $\theta$ near $\theta_{m}$, becomes very large as the ray approaches the saddle point.  As the expanded variables become very large, computing $Z$ using equation (\ref{eq:56}) is unstable.  Furthermore, this effect becomes worse as the initial value, $\omega = \Phi(0)$, goes to zero.

\section*{Acknowledgments} 
JMN would like to thank James P. Keener for valuable discussions throughout this project.  This work was supported by Award No KUK-C1-013-4 made by King Abdullah University of Science and Technology (KAUST).

\end{document}